\documentstyle[11pt]{article}
\newcounter{subequation}[equation]

\makeatletter
\def\thesubequation{\theequation\@alph\c@subequation}
\def\@subeqnnum{{\rm (\thesubequation)}}
\def\slabel#1{\@bsphack\if@filesw {\let\thepage\relax
   \xdef\@gtempa{\write\@auxout{\string
      \newlabel{#1}{{\thesubequation}{\thepage}}}}}\@gtempa
   \if@nobreak \ifvmode\nobreak\fi\fi\fi\@esphack}
\def\subeqnarray{\stepcounter{equation}
\let\@currentlabel=\theequation\global\c@subequation\@ne
\global\@eqnswtrue
\global\@eqcnt\z@\tabskip\@centering\let\\=\@subeqncr
$$\halign to \displaywidth\bgroup\@eqnsel\hskip\@centering
  $\displaystyle\tabskip\z@{##}$&\global\@eqcnt\@ne
  \hskip 2\arraycolsep \hfil${##}$\hfil
  &\global\@eqcnt\tw@ \hskip 2\arraycolsep
  $\displaystyle\tabskip\z@{##}$\hfil
   \tabskip\@centering&\llap{##}\tabskip\z@\cr}
\def\endsubeqnarray{\@@subeqncr\egroup
                     $$\global\@ignoretrue}
\def\@subeqncr{{\ifnum0=`}\fi\@ifstar{\global\@eqpen\@M
    \@ysubeqncr}{\global\@eqpen\interdisplaylinepenalty \@ysubeqncr}}
\def\@ysubeqncr{\@ifnextchar [{\@xsubeqncr}{\@xsubeqncr[\z@]}}
\def\@xsubeqncr[#1]{\ifnum0=`{\fi}\@@subeqncr
   \noalign{\penalty\@eqpen\vskip\jot\vskip #1\relax}}
\def\@@subeqncr{\let\@tempa\relax
    \ifcase\@eqcnt \def\@tempa{& & &}\or \def\@tempa{& &}
      \else \def\@tempa{&}\fi
     \@tempa \if@eqnsw\@subeqnnum\refstepcounter{subequation}\fi
     \global\@eqnswtrue\global\@eqcnt\z@\cr}
\let\@ssubeqncr=\@subeqncr
\@namedef{subeqnarray*}{\def\@subeqncr{\nonumber\@ssubeqncr}\subeqnarray}
\@namedef{endsubeqnarray*}{\global\advance\c@equation\m@ne%
                           \nonumber\endsubeqnarray}

\makeatletter
\@addtoreset{equation}{section}
\makeatother
\renewcommand{\theequation}{\thesection.\arabic{equation}}
\def\dalemb#1#2{{\vbox{\hrule height .#2pt
        \hbox{\vrule width.#2pt height#1pt \kern#1pt
                \vrule width.#2pt}
        \hrule height.#2pt}}}

    \let\e=\epsilon
  \let\q=\theta  
  \let\n=\nu

\def\nn{\nonumber} \def\bd{\begin{document}} \def\ed{\end{document}}
\def\ds{\documentstyle} \let\fr=\frac \let\bl=\bigl \let\br=\bigr
\let\Br=\Bigr \let\Bl=\Bigl
\let\bm=\bibitem
\let\na=\nabla
\let\pa=\partial \let\ov=\overline
\def\ie{{\it i.e.\ }}
\newcommand{\be}{\begin{equation}}
\newcommand{\ee}{\end{equation}}
\def\ba{\begin{array}}
\def\ea{\end{array}}
\def\ft#1#2{{\textstyle{{\scriptstyle #1}\over {\scriptstyle #2}}}}
\def\fft#1#2{{#1 \over #2}}
\def\del{\partial}
\def\sst#1{{\scriptscriptstyle #1}}
\def\oneone{\rlap 1\mkern4mu{\rm l}}
\def\e7{E_{7(+7)}}
\def\td{\tilde}
\def\wtd{\widetilde}
\def\im{{\rm i}}
\def\bog{Bogomol'nyi\ }
\def\q{{\tilde q}}
\def\hast{{\hat\ast}}
\def\0{{\sst{(0)}}}
\def\1{{\sst{(1)}}}
\def\2{{\sst{(2)}}}
\def\3{{\sst{(3)}}}
\def\4{{\sst{(4)}}}
\def\5{{\sst{(5)}}}
\def\6{{\sst{(6)}}}
\def\7{{\sst{(7)}}}
\def\8{{\sst{(8)}}}
\def\n{{\sst{(n)}}}

\def\hA{\hat{\cal A}}
\def\ns{{\sst {\rm NS}}}
\def\rr{{\sst {\rm RR}}}
\def\tH{{\widetilde H}}
\def\tB{{\widetilde B}}
\def\cA{{\cal A}}
\def\cF{{\cal F}}
\def\tF{{\wtd F}}
\def\v{{{\cal V}}}
\def\Z{\rlap{\sf Z}\mkern3mu{\sf Z}}
\def\ep{{\epsilon}}
\def\IIA{{\rm IIA}}
\def\IIB{{\rm IIB}}
\def\ads{{\rm AdS}}
\def\R{\rlap{\rm I}\mkern3mu{\rm R}}

\newcommand{\ho}[1]{$\, ^{#1}$}
\newcommand{\hoch}[1]{$\, ^{#1}$}
\newcommand{\bea}{\begin{eqnarray}}
\newcommand{\eea}{\end{eqnarray}}
\newcommand{\ra}{\rightarrow}
\newcommand{\lra}{\longrightarrow}
\newcommand{\Lra}{\Leftrightarrow}
\newcommand{\ap}{\alpha^\prime}
\newcommand{\bp}{\tilde \beta^\prime}
\newcommand{\tr}{{\rm tr} }
\newcommand{\Tr}{{\rm Tr} }
\newcommand{\NP}{Nucl. Phys. }
\newcommand{\tamphys}{\it Center for Theoretical Physics,
Texas A\&M University, College Station, Texas 77843}
\newcommand{\ens}{\it Laboratoire de Physique Th\'eorique de l'\'Ecole
Normale Sup\'erieure\hoch{3,4}\\
24 Rue Lhomond - 75231 Paris CEDEX 05}

\newcommand{\auth}{M.J. Duff\hoch{\ddagger1}, H. L\"u\hoch{\dagger} and
C.N. Pope\hoch{\ddagger\star2}}

\begin{document}
\begin{flushright}
\hfill{CTP TAMU-28/98, \ \ LPTENS-98/30, \ \ SISSARef. 79/98/EP, \ \ July
  1998}\\
\hfill{\bf hep-th/9807173}\\
\end{flushright}

%\vspace{10pt}

\begin{center}
{\large {\bf AdS$_3\times S^3$ (Un)twisted and Squashed, and \\
   An $O(2,2;\Z)$ Multiplet of Dyonic Strings}}

\vspace{20pt}

\auth

\vspace{10pt}
{\hoch{\dagger}\ens}

%\vspace{10pt}
{\hoch{\ddagger}\tamphys}

{\hoch{\star}{SISSA, Via Beirut No. 2-4, 34013 Trieste, Italy\hoch{3}}}

\vspace{20pt}

\underline{ABSTRACT}
\end{center}

We consider type IIB configurations carrying both NS-NS and R-R
electric and magnetic 3-form charges, and whose near horizon geometry
contains AdS$_{3} \times S^{3}$.  Noting that $S^{3}$ is a $U(1)$
bundle over $CP^{1} \sim S^{2}$, we construct the dual type IIA
configurations by a Hopf T-duality along the $U(1)$ fibre. In the case
where there are only R-R charges, the $S^{3}$ is untwisted to
$S^{2}\times S^{1}$ (in analogy with a previous treatment of AdS$_{5}
\times S^{5}$).  However, in the case where there are only NS-NS
charges, the $S^{3}$ becomes the cyclic lens space $S^{3}/Z_{p}$ with
its round metric (and is hence invariant when $p=1$), where $p$ is the
magnetic NS-NS charge.  In the generic case with NS-NS and R-R
charges, the $S^{3}$ not only becomes $S^{3}/Z_{p}$ but is also
squashed, with a squashing parameter that is related to the values of
the charges.  Similar results apply if we regard AdS$_{3}$ as a bundle
over AdS$_{2}$ and T-dualise along the fibre.  We show that Hopf
T-dualities relate different black holes, and that they preserve the
entropy.  The AdS$_3\times S^3$ solutions arise as the near-horizon
limits of dyonic strings. We construct an $O(2,2;\Z)$ multiplet of
such dyonic strings, where $O(2,2;\Z)$ is a subgroup of the $O(5,5)$
or $O(5,21)$ six-dimensional duality groups, which captures the
essence of the NS-NS/R-R and electric/magnetic dualities.

{\vfill\leftline{}\vfill
\vskip  10pt
\footnoterule
{\footnotesize
        \hoch{1} Research supported in part by NSF Grant PHY-9722090.
\vskip  -12pt} \vskip   14pt
{\footnotesize
        \hoch{2}        Research supported in part by DOE
grant DE-FG03-95ER40917. \vskip -12pt}  \vskip  14pt
{\footnotesize \hoch{3} Supported in part by EC under TMR
contract ERBFMRX-CT96-0045 \vskip -12pt} \vskip 14pt
{\footnotesize
        \hoch{4} Unit\'e Propre du Centre National de la Recherche
Scientifique, associ\'ee \`a l'\'Ecole Normale Sup\'erieure
\vskip -12pt} \vskip 10pt
{\footnotesize \hoch{\phantom{3}} et \`a l'Universit\'e de Paris-Sud.
\vskip -12pt} \vskip 10pt}

\pagebreak
\setcounter{page}{1}

\tableofcontents
\addtocontents{toc}{\protect\setcounter{tocdepth}{2}}
\newpage

\section{Introduction\label{intro}}

The six-dimensional space AdS$_{3}\times S^{3}$ emerges as the near
horizon geometry \cite{DGT,GHT} of the self-dual string
\cite{lublack,DKL} or, more generally, the dyonic string
\cite{Rahmfeld,DKL,DLPhet,DLPgauge}. The dyonic string admits the
ten-dimensional interpretation \cite{Rahmfeld} of an intersecting
NS-NS 1-brane and 5-brane, which in a type II context is in turn
related by U-duality to the D1-D5 brane system
\cite{Boonstra1,Boonstra2,Horowitz2,Horowitz3,Maldacena3}. This
geometry plays a part in recent
studies of black holes \cite{callan,Tseytlin1,Cvetic1,Cvetic2,lpmult,Vafa,%
  Horowitz2,Horowitz3,Maldacena3,DBISSS} and has attracted a good deal
of attention lately
\cite{Maldacena,behrndt,evans,andrianopoli,Maldacena3,navarro,lee,%
andreas,deger,benakli,Boer,Vafa2,Boonstra,CKKTV,Ortiz,giveon,martinec,nk2}
following the conjectured duality \cite{Maldacena} between physics in
the bulk of the anti-de Sitter spacetimes and conformal field theories
on their boundary. AdS$_{3}$ is particularly interesting in this
regard because the conformal field theory is then of the familiar and
well-understood $1+1$ dimensional variety.

In a previous paper \cite{DLP2}, devoted mainly to AdS$_{5} \times
S^{5}$, we noted that odd-dimensional spheres $S^{2n+1}$ may be
regarded as $U(1)$ bundles over $CP^{n}$ and that this permits an
unconventional type of T-duality along the $U(1)$ fibre that we called
``Hopf'' duality. This Hopf duality has the effect of untwisting
$S^{2n+1}$ to $CP^{n} \times S^{1}$.  Applying this to $S^{5}$, we
were able to construct the duality chain: $n=4$ Yang-Mills
$\rightarrow$ type IIB string on AdS$_{5} \times S^{5}$ $\rightarrow$
type IIA string on AdS$_{5} \times CP^{2} \times S^{1}$ $\rightarrow$
M-theory on AdS$_{5} \times CP^{2} \times T^{2}$. In an earlier paper
\cite{swos}, devoted mainly to AdS$_{4} \times
S^{7}$, we exhibited the duality: M-theory on AdS$_{4} \times
S^{7}$ $\rightarrow$
type IIA string on AdS$_{4} \times CP^{3}$. In both
contexts, similar techniques
were also applied to more general spaces AdS$ \times M $ where $M$ are
Einstein spaces that are not necessarily spheres.  These emerge as the
near horizon geometries of supermembranes with fewer Killing spinors
\cite{DLPS,ferrara,ccdfft} and whose boundary conformal field theories have
correspondingly less supersymmetry \cite{kachru,DLP2,ferrara,Oz,klebanov}.

In the present paper, we wish to apply these techniques to find type
IIA and M-theory duals of six-dimensional type IIB AdS$_{3}\times
S^{3}$ configurations obtained by either $T^{4}$ or K3
compactifications. The novel ingredient is that these can be supported
by both NS-NS and R-R 3-forms, in contrast to the AdS$_{5} \times
S^{5}$ example where the 5-form was strictly R-R. This has some
interesting and unexpected consequences.  Noting that $S^{3}$ is a
$U(1)$ bundle over $CP^{1} \sim S^{2}$, we construct the dual type IIA
configurations by a Hopf T-duality along the $U(1)$ fibre. In the case
where there are only R-R charges, the $S^{3}$ is untwisted to
$S^{2}\times S^{1}$ (in analogy with a previous treatment of AdS$_{5}
\times S^{5}$).  However, in the case where there are only NS-NS
charges, the $S^{3}$ becomes the cyclic lens space $S^{3}/Z_{p}$ with
its round metric (and is hence invariant when $p=1$), where $p$ is the
magnetic NS-NS charge.  In the generic case with NS-NS and R-R
charges, the $S^{3}$ not only becomes $S^{3}/Z_{p}$ but is also
squashed, with a squashing parameter that is related to the values of
the charges.  Similar results apply if we regard AdS$_{3}$ as a bundle
over AdS$_{2}$ and T-dualise along the fibre.  We note that these Hopf
dualities preserve the area of the horizons, and hence they preserve
the black hole entropies.  In particular, we show in appendix D that
a Hopf reduction of a single 3-charge isotropic black hole in $D=5$
gives a 4-charge black hole in $D=4$, where the fourth charge is of
unit strength, and comes from the magnetic charge carried by the
Kaluza-Klein vector.  The reduction coordinate in this case is the
$U(1)$ fibre coordinate of the foliating 3-spheres in the transverse
space of the $D=5$ black hole.  In general Hopf reduction and Hopf
T-duality not only relate the near-horizon limits of black holes, but
also relate the full solutions themselves.  We show also that this
statement extends to all $p$-branes, including non-extremal ones, that
have 4-dimensional overall transverse spaces.  In all cases, the Hopf
reduction maps an $N$-charge solution in $D+1$ dimensions to an
$(N+1)$-charge solution in $D$ dimensions.

Studying the T-duality and the U-duality multiplets of BPS solitons in
the full theories obtained from the $T^4$ or K3 reductions of
ten-dimensional supergravities is a complicated matter, owing to the
large number of fields, and the size of the global symmetry groups in
$D=6$.  For this reason, it is helpful to make truncations of the
six-dimensional theories, to more manageable subsectors that capture
the essential features that we wish to study.  We therefore begin, in
section 2, by making a consistent truncation of six-dimensional
maximal supergravity, to a subsector of bosonic fields that includes
two 2-form potentials, one NS-NS, and the other R-R.  We show that
this theory has an $O(2,2)\sim SL(2,\R)_1 \times SL(2,\R)_2$ global
symmetry.  The $SL(2,\R)_1$ describes an S-duality symmetry that
interchanges the NS-NS and R-R 2-form potentials, while the
$SL(2,\R)_2$ is an electric/magnetic duality symmetry of the 3-form
field strengths, which acts locally only at the level of the equations
of motion.  This consistent truncation is most conveniently
constructed from the toroidal reduction in the type IIB field
variables.  We then consider a different consistent truncation, which
is most conveniently obtained from the toroidal reduction in the type
IIA fields variables.  In fact the two truncations are characterised
by the feature that the six-dimensional fields that are retained are
precisely the original ten-dimensional ones, with the spacetime
indices simply restricted to run over the six-dimensional range, and
in addition the breathing-mode scalar parameterising the volume of the
4-torus.  For this reason, the truncated theories can equally well be
obtained by compactifying the type IIB and type IIA theories on K3,
and following the identical prescription for which fields to retain.

The two truncated theories in $D=6$ are related by a T-duality
transformation upon reduction on $S^1$ to $D=5$.  In section 2,
supplemented by appendix A, we obtain the above truncations and
perform the $S^1$ reductions on the two theories.  We then derive the
explicit T-duality transformations relating them.  In appendix B, we
study the $O(2,2)$ symmetry in detail, and give the explicit
transformation rules.  In appendix C, we use these transformation
rules to construct an $O(2,2;\Z)$ multiplet of dyonic strings carrying
four independent charges, namely electric and magnetic charges for
each of the NS-NS and R-R 3-forms.  We then extend these results to
boosted and twisted dyonic strings. Since all the solutions are obtained as
solutions in a consistent truncation, the dyonic strings are therefore
solutions of the original supersymmetric theories, and in fact they
are, as usual, BPS states preserving some fraction of the
supersymmetry.

In section 3, we study the near-horizon AdS$_3\times S^3$ limits of
the dyonic strings.  Noting that $S^3$ can be described as a $U(1)$
bundle over $S^2$, we perform a Hopf T-duality transformation on the
$U(1)$ fibres, and show that the $S^3$ can be untwisted or squashed,
as described previously.  In the case of solutions supported purely by
NS-NS fields, we also supply a CFT proof that strings on $S^3/Z_m$
with 3-form flux $n$ are dual to strings on $S^3/Z_n$ with 3-form flux
$m$.  

     In section 4, we perform a similar Hopf
T-duality transformation on the AdS$_3$ instead, exploiting the fact
that AdS$_3$ can analogously be written in the form of a bundle over
AdS$_2$.  In section 5 we perform simultaneous Hopf T-duality
transformations on the fibres of $S^3$ and AdS$_3$.  Section 6
addresses the issue of supersymmetry and the Hopf T-duality
transformations.  We construct the Killing spinors on AdS$_3$ and
$S^3$ explicitly, in coordinates appropriate to the bundle
descriptions, and show that Hopf T-duality on AdS$_3$ or $S^3$ either
preserves all or none of the supersymmetry, at the level of the
massless Kaluza-Klein modes in supergravity, depending on the
orientation of the fibration.  We also discuss the supersymmetry in
the context of the full string theory.

   In section 7, we list all the non-dilatonic black holes in $D=5$
and $D=4$, and study their near-horizon limits when they are oxidised
to $D=6$.  We show that all the near-horizon limits can be obtained by
Hopf T-duality on AdS$_3\times S^3$.

\section{$O(2,2)$ truncation of maximal supergravity in $D=6$}

We begin from the Lagrangian in $D=6$ obtained by dimensional
reduction of type IIB on a 4-torus.  Since we want to consider
AdS$_3\times S^3$ solutions that carry both NS-NS and R-R 3-form
charges, we first make a consistent truncation to a subset of the
fields that includes the necessary pair of 3-forms.  We can do this by
just retaining the subset of fields comprising the reductions of the
original $F_\3^\ns$ and $F_\3^\rr$ fields themselves, together with
the axion $\chi_1$ and the dilaton $\hat\phi$ of $D=10$ type IIB, and
the axion $\chi_2$ coming from the dualisation of the potential
$B_\4$.  (All details of the full reduction are given in \cite{llpt}.)
In $D=6$, we obtain
%%%%%
\bea
e^{-1}\, {\cal L}_6 &=& R -\ft12 (\del\hat\phi)^2
-\ft12(\del\vec\varphi)^2 -\ft12 e^{2\hat\phi}\, (\del\chi_1)^2
-\ft12 e^{-2\vec a\cdot\vec\varphi}\, (\del \chi_2)^2 \nn\\
&&-\ft1{12} e^{-\hat\phi +\vec a\cdot\vec\varphi}\, (F_\3^\ns)^2
-\ft1{12} e^{\hat\phi +\vec a\cdot\vec\varphi}\, (F_\3^\rr)^2
+ \chi_2\, dA_\2^\ns\wedge dA_\2^\rr\ ,
\eea
%%%%%
where $F_\3^\ns=dA_\2^\ns$ and $F_\3^\rr=dA_\2^\rr +\chi_1\,
dA_\2^\ns$. Here $\vec\varphi$ denotes the set of 4 dilatonic scalars
coming from the reduction on $T^4$, and $\vec a$ is a constant vector that
can be found in \cite{llpt}, characterising the couplings of the
field strengths to the dilatonic scalars $\vec\varphi$.   The
combinations of the dilatons $\vec\varphi$ that are perpendicular to
$\vec a$ can also be consistently truncated, resulting in the
Lagrangian
%%%%%
\bea
e^{-1}\, {\cal L}_{6B} &=& R -\ft12 (\del\phi_1)^2
-\ft12(\del\phi_2)^2 -\ft12 e^{2\phi_1}\, (\del\chi_1)^2
-\ft12 e^{2\phi_2}\, (\del \chi_2)^2 \nn\\
&&-\ft1{12} e^{-\phi_1 -\phi_2}\, (F_\3^\ns)^2
-\ft1{12} e^{\phi_1 -\phi_2}\, (F_\3^\rr)^2
+ \chi_2\, dA_\2^\ns\wedge dA_\2^\rr\ ,\label{2blag}
\eea
%%%%%
where $\phi_1=\phi$ and $\phi_2= -\vec a\cdot\vec \varphi$.
This truncated Lagrangian is characterised by the fact that it follows
from the truncation of the $T^4$ reduction of the type IIB theory in
which all the original fields are retained, but with their indices now
running only over the remaining six dimensions.  (Note that the
potential for the self-dual 5-form is now dualised to give the axion
$\chi_2$.)  In addition, the breathing-mode scalar $\phi_2$ that
parameterises the volume of $T^4$ is also included, but all other
fields with indices internal to the 4-torus are set to zero.  For this
reason, (\ref{2blag}) can also be obtained by making a consistent
truncation of the $N=2$ theory obtained by compactifying type IIB on
K3, following the identical prescription for which fields are to be
retained.  (One cannot tell what it is that is ``breathing'' if all
other modes are truncated.)

The Lagrangian (\ref{2blag}) has an $O(2,2)\sim SL(2,\R)_1\times
SL(2,\R)_2$ global symmetry which is a subgroup of the original
$O(5,5)$ Cremmer-Julia symmetry of maximal supergravity in $D=6$.
Note that $SL(2,\R)_1$, realised by $(\phi_1,\chi_1)$ in the scalar
sector, is a symmetry of the Lagrangian, whilst $SL(2,\R)_2$, realised
by $(\phi_2,\chi_2)$, is a symmetry of the equations of motion.  We
give the explicit transformation rules in appendix B. It should
emphasised that this $O(2,2)$-invariant theory is not itself the
bosonic sector of any supergravity theory; rather, it is a convenient
{\it consistent} truncation of $D=6$ maximal supergravity that
contains all the fields necessary for describing the (supersymmetric)
solutions of $D=6$ maximal supergravity that we are going to discuss
in this paper.\footnote{A similar use of a non-supersymmetric
  truncation was made in \cite{schwarz}, where the 5-form field
  strength of the type IIB theory was set to zero in order to simplify
  the discussion of the multiplet of supersymmetric NS-NS and R-R
  string solitons.  In that case the truncation was actually {\it
    inconsistent}, since generic solutions of the truncated theory
  would be configurations which, in the full type IIB theory, would
  provide sources that would force the 5-form field strength to be
  non-zero.  However, these source terms are actually zero for the
  class of solutions considered in \cite{schwarz}, and so the
  truncation there was consistent only in this restricted sense.}
Because of the consistency of the truncation, all the solutions in the
theory (\ref{2blag}) are solutions of the original untruncated
supersymmetric theory.  This provides a powerful tool for studying the
BPS states of the original theory, since it reduces the original
$O(5,5)$ or $O(5,21)$ global symmetry to a more manageable $O(2,2)$
global symmetry that nevertheless captures the essence of the
NS-NS/R-R and the electric/magnetic dualities.

The six-dimensional Lagrangian (\ref{2blag}) is related by T-duality
in $D=5$ to a different six-dimensional theory that is also a
consistent truncation of maximal six-dimensional supergravity.  This
theory is most conveniently obtained by truncating maximal
six-dimensional supergravity described in terms of the $T^4$ reduction
in the type IIA field variables.  In this description, it corresponds
again to retaining the original ten-dimensional fields, with their
indices running only over the remaining six dimensions, and including
in addition the breathing-mode scalar parameterising the volume of the
4-torus. Again, as with (\ref{2blag}), all other six-dimensional
fields, with indices internal to the 4-torus, are set to zero.  This
truncated Lagrangian can therefore also be obtained as a consistent
truncation of the $N=2$ theory obtained by K3 compactification of type
IIA, where the same truncation prescription is applied.  It is given
by
%%%%%
\bea
e^{-1}\, {\cal L}_{6A} &=& R -\ft12 (\del\phi_1)^2
-\ft12(\del\phi_2)^2 -\ft1{48} e^{\fft12\phi_1-\fft32\phi_2}
\, (F_\4)^2\nn\\
&&-\ft1{12} e^{-\phi_1-\phi_2}\, (F_\3)^2
-\ft1{4} e^{\fft32\phi_1 -\fft12\phi_2}\, (F_\2)^2
\ ,\label{2alag}
\eea
%%%%%
where, in the notation of \cite{lpsol,cjlp1}, $F_\3$ means the NS-NS
3-form $F_{\3 1}$, and $F_\2$ means the R-R 2-form ${\cal F}_\2^1$,
with the index ``1'' here denoting the reduction step from $D=11$ to
$D=10$.  (In the rest of the paper, an index ``1'' will be used
exclusively to denote a reduction step from 6 to 5 dimensions.)  We
have again performed a consistent truncation and orthogonal
transformation on the dilatons.  To be precise, $\phi_1$ is the
original ten-dimensional dilaton, and $\phi_2$ is the breathing mode
of the 4-torus or K3.  Note that $F_\4=dA_\3 -dA_\2\wedge A_\1$, while
$F_\3=dA_\2$ and $F_\2=dA_\1$.  The $D=6$ string coupling constant in
both (\ref{2blag}) and (\ref{2alag}) is given by $g_6 =
e^{\fft12(\phi_1+\phi_2)}$.  The Lagrangian (\ref{2alag}) is in fact
simply the dimensional reduction of
%%%%%
\be
e^{-1}\, {\cal L}_7 = R -\ft12 (\del\phi)^2 -\ft1{48}
e^{\sqrt{\ft85}\phi}\, F_\4^2\ .\label{7dlag}
\ee
%%%%%
This provides another way of verifying a statement we made previously,
that the six-dimensional Lagrangians (\ref{2alag}) and (\ref{2blag})
are consistent truncations of six-dimensional maximal supergravity.
We do this by noting that (\ref{7dlag}) is itself a consistent
truncation of seven-dimensional maximal supergravity.  This can be
easily seen from the form of the $D=7$ theory given in \cite{lpsol},
and by noting that $F_\4$ cannot act as a source for any of the other
fields that are being truncated.  (Recall that for a truncation to be
consistent, every solution of the truncated theory must be a solution
of the untruncated theory.)  Having established that (\ref{7dlag})
is a consistent truncation of seven-dimensional maximal supergravity,
it follows that (\ref{2alag}) and (\ref{2blag}) are consistent
truncations of six-dimensional maximal supergravity. (For the latter
case, one has to invoke the T-duality relating the two theories in
$D=5$.)

To see the T-duality relating (\ref{2blag}) and (\ref{2alag}), we
dimensionally reduce the two six-dimensional theories to $D=5$; the
details are given in appendix A.  The T-duality relations between the
field strengths in the five-dimensional theories are indicated in
Table 1 below, which also defines our notation for the dimensional
reductions of the fields.  Note that we present the identifications at
the level of the field strengths because it is necessary to perform
some dualisations in $D=5$ in order to implement the
identifications.\footnote{Note that perturbative T-duality does not
  require dualisations in order to relate two theories.  The reason
  why it is necessary to make a dualisation here is that we have
  already dualised the 4-form potential in the type IIB picture to an
  axion.  Had we not done so, then the identification of
  five-dimensional fields would not have required any dualisation.}
(The precise statement of the identifications is given in appendix A.)
Note also that we are now, and henceforth, using a ``1'' subscript on
a field strength to denote the reduction step from $D=6$ to $D=5$.

\bigskip\bigskip
\begin{center}
\begin{tabular}{|c|c|c|c|c|c|}\hline
    &\multicolumn{2}{|c|}{IIA} &
    &\multicolumn{2}{c|}{IIB} \\ \cline{2-6}
    & $D=6$ & $D=5$ &T-duality & $D=5$ & $D=6$ \\ \hline\hline
    & $F_\4$ & $F_\4$ & $\longleftrightarrow$ &
                   $d\chi_2$ & $d\chi_2$ \\ \cline{3-6}
R-R & &  ${F_{\3 1}}$& $\longleftrightarrow$
                           & $F_\3^{\rr}$ & $F_\3^{\rr}$
                                               \\ \cline{2-5}
fields& $F_\2$ & $F_\2$ &
                $\longleftrightarrow$ &
        ${F_{\2 1}^{\rr}}$ & \\ \cline{3-6}
   & & $F_{\1 1}$ & $\longleftrightarrow$
                            & $d\chi_1$ &$d\chi_1$
                                 \\ \hline\hline
NS-NS & $G_{\mu\nu}$ & ${{\cal F}_\2}$
                        & $\longleftrightarrow$ &
        ${F_{\2 1}^{\ns}}$ & $F_\3^{\ns}$ \\ \cline{2-5}
fields& $F_{\3}$ & $F_{\3 }$ &
               $\longleftrightarrow$ & $F_\3^{\ns}$ &
                                      \\ \cline{3-6}
      & & ${F_{\2 1}}$ & $\longleftrightarrow$ &
                              ${{\cal F}_\2}$ & $G_{\mu\nu}$
                                       \\ \hline
\end{tabular}
\end{center}

\bigskip

\centerline{Table 1: Fields of the truncated type II
theories in $D=6$ and $D=5$}
\bigskip\bigskip

   Our ansatz for the reduction of the metric from $D=6$ to $D=5$ is
%%%%%%%%%%%%%
\be
ds_6^2 = e^{-\varphi/\sqrt6}\, ds_5^2 + e^{\sqrt{\fft32}\varphi}\,
(dz + \cA_\1)^2\ .\label{metred}
\ee
%%%%%%%%
We find that the dimensional reductions of the two Lagrangians
(\ref{2blag}) and (\ref{2alag}) become equivalent, after making the
identifications given in Table 1, provided that the dilatons of the
two theories are related by the orthogonal transformation
%%%%%%%%%
\be
\pmatrix{\phi_1\cr \phi_2\cr \varphi}_{\sst\rm IIA} =
\Lambda\, \pmatrix{\phi_1\cr \phi_2\cr \varphi}_{\sst\rm IIB}=
 \pmatrix{\ft34 & -\ft14 & -\sqrt{\ft38} \cr
 -\ft14 & \ft34 & -\sqrt{\ft38} \cr
 -\sqrt{\ft38} & -\sqrt{\ft38} & -\ft12}\,
\pmatrix{\phi_1\cr \phi_2\cr \varphi}_{\sst\rm IIB}\ .\label{lambda}
\ee
%%%%%%%%%%
Note that this matrix satisfies $\Lambda=\Lambda^{-1}$.  In terms of
the string metrics, the radius of the compactifying circle is given by
$R=e^{\fft14\phi_1 +\fft14 \phi_2 + \sqrt{\ft38}\varphi}$.  It is
easily seen that under the transformation (\ref{lambda}) we have
$R_{\sst\rm IIA} = 1/R_{\sst\rm IIB}$.

     It is worth remarking that the Lagrangians (\ref{2blag}) and
(\ref{2alag}) are both consistent truncations of maximal
supergravity.  It follows that their respective solutions that are
related by T-duality transformation are all solutions of the
untruncated maximal supergravity.  If instead we consider the two
Lagrangians (\ref{2blag}) and (\ref{2alag}) as consistent truncations
of the K3 compactifications of the type IIB and type IIA
supergravities, then their solutions that are related by T-duality
remain as distinct solutions of the original type IIB and type IIA
supergravities.

    In the rest of the paper, we shall refer to the two Lagrangians
(\ref{2blag}) and (\ref{2alag}) as the type IIB and the type IIA
descriptions of the six-dimensional  truncated theories.

\section{$S^3$ (un)twisted and squashed}

In the previous section, we obtained truncated six-dimensional type
IIB and type IIA Lagrangians, and their T-duality relation.  We are
now in a position to consider the AdS$_3\times S^3$ solution.  The
Lagrangian (\ref{2blag}) admits dyonic string solutions supported
either by the NS-NS 3-form $F_\3^\ns$ or the R-R 3-form $F_\3^\rr$.
More general solutions can be obtained by acting with the $O(2,2)$
symmetry of the theory, allowing us, in particular, to find solutions
for dyonic strings carrying both NS-NS and R-R charges.  We do this in
detail in appendix C, obtaining an $O(2,2;\Z)$ multiplet of dyonic
strings.

Near the horizon, even though the above dyonic solutions carry four
independent charges, the 3-forms $F_\3^\ns$ and $F_\3^\rr$ become
self-dual, and the metric approaches that of AdS$_3\times S^3$.  In
fact it is more convenient to construct these solutions directly.  The
dilatons $\phi_1$ and $\phi_2$ and the axions $\chi_1$ and $\chi_2$
are constant in the solution, and for simplicity we shall take them to
be zero.  The remaining equations are solved by taking the metric and
3-forms to be
%%%%%
\bea
ds_6^2 &=& ds^2(\ads) + ds^2(S^3)\ ,\nn\\
F_\3^\ns &=& \lambda\, \ep(\ads) + \lambda\, \ep(S^3)\ ,\label{adssol}\\
F_\3^\rr &=& \mu\, \ep(\ads) + \mu\, \ep(S^3)\ ,\nn
\eea
%%%%%
where $\lambda$ and $\mu$ are constants, and the metrics on the
AdS$_3$ and $S^3$ have Ricci tensors given by
%%%%%
\be
R_{\mu\nu} = -\ft12(\lambda^2 +\mu^2)\, g_{\mu\nu}\ ,\qquad
R_{mn} = \ft12(\lambda^2 +\mu^2)\, g_{mn}\label{ricci}
\ee
%%%%%
respectively.  (It is easy to see from (\ref{2blag}) that the
equations of motion will only be satisfied by taking the dilatons to
be constant if the coefficients in front of the volume forms
$\ep(\ads)$ and $\ep(S^3)$ for the AdS$_3$ and $S^3$ metrics are
equal, and hence the 3-form field strengths are self-dual.)  The
constants $\lambda$ and $\mu$ are related to the magnetic charges as
follows:
%%%%%
\be
Q_\ns \equiv \ft1{16\pi^2}
        \int F_\3^\ns = \fft{\lambda}{(\lambda^2 +\mu^2)^{3/2}}\ ,
\qquad
Q_\rr \equiv \ft1{16\pi^2}
     \int F_\3^\rr = \fft{\mu}{(\lambda^2 +\mu^2)^{3/2}}
\ .\label{charges}
\ee
%%%%%
In calculating the charges, we have made use of the fact that a
3-sphere whose Ricci tensor is given by $R_{mn}$ in (\ref{ricci}) has
volume $16\pi^2 (\lambda^2+\mu^2)^{-3/2}$, and in fact its metric can
be written as
%%%%%
\be
ds^2(S^3) = \fft4{\lambda^2 + \mu^2}\, d\Omega_3^2\ ,
\ee
%%%%%
where $d\Omega_3^2$ is the metric on a unit 3-sphere.  Note that since
we are taking the constant values of the two dilatons $\phi_1$ and
$\phi_2$ to be zero for simplicity, this means that the electric
charges (whose calculation we have not explicitly presented above) are
equal to the magnetic charges.  If one chooses non-zero values for the
constants $\phi_1$ and $\phi_2$ then the field strengths are still
self-dual, but the electric and magnetic charges will be unequal.
Note that the charges will all be integers.

   We now make use of the fact that the metric $d\Omega_3^2$ can be
written as a $U(1)$ bundle over $CP^1\sim S^2$ as follows:
%%%%%
\be
d\Omega_3^2 = \ft14 d\Omega_2^2 +\ft14 (dz + B)^2 \ ,\label{fib32}
\ee
%%%%%
where $d\Omega_2^2$ is the metric on the unit 2-sphere, whose volume
form $\Omega_\2$ is given by $\Omega_\2= dB$.  (If $d\Omega_2^2$ is
written in spherical polar coordinates as $d\Omega_2^2 = d\theta^2 +
\sin^2\theta\, d\phi^2$, then we can write $B$ as $B=\cos\theta\,
d\phi$.) The fibre
coordinate $z$ has period $4\pi$.  Thus the six-dimensional metric
given in (\ref{adssol}) can be written as
%%%%%
\be
ds_6^2 = ds^2(\ads) + \fft1{\lambda^2+\mu^2}\, d\Omega_2^2 +
\fft1{\lambda^2+\mu^2}\, (dz + B)^2 \ .\label{2bmet}
\ee
%%%%%
The four-dimensional area of the horizon is given by
%%%%%%%%
\be A\sim L\, (\lambda^2+\mu^2)^{-3/2}\ ,\label{area} \ee
%%%%%%%%
where $L$ is the contribution from $ds^2({\rm AdS})$ at the boundary at
constant time.  The field strengths in (\ref{adssol}) can now be written as
%%%%%
\bea
F_\3^\ns &=& \lambda\, \ep(\ads) +
\fft{\lambda}{(\lambda^2+\mu^2)^{3/2}}\, \Omega_\2\wedge(dz +B)
\ ,\nn\\
F_\3^\rr &=& \mu\, \ep(\ads) +
\fft{\mu}{(\lambda^2+\mu^2)^{3/2}}\, \Omega_\2\wedge(dz + B)\ .
\label{fnfs}
\eea
Comparing (\ref{2bmet}) with the general reduction ansatz
(\ref{metred}), we see that if we dimensionally reduced on the fibre
coordinate we obtain the 5-dimensional metric
%%%%%
\be
ds_5^2 = (\lambda^2+\mu^2)^{-1/3}\, ds^2(\ads) +
(\lambda^2+\mu^2)^{-4/3}\, d\Omega_2^2\ ,
\ee
while the new dilaton $\varphi$ is a constant, given by
\be
e^{\varphi/\sqrt6} = (\lambda^2+\mu^2)^{-1/3}\ .\label{varphi}
\ee
%%%%%
Comparing (\ref{fnfs}) with the reduction ans\"atze $F_\n \rightarrow
F_\n + F_{\sst{(n-1)}}\wedge (dz+B)$ for the field
strengths, we find that in $D=5$ we have
%%%%%
\bea
&&F_\3^\ns = \lambda\, \ep(\ads)\ ,\qquad
F_{\2 1}^\ns = \fft{\lambda}{(\lambda^2+\mu^2)^{3/2}}\, \Omega_\2
\ ,\nn\\
&&F_\3^\rr = \mu\, \ep(\ads)\ ,\qquad
F_{\2 1}^\rr = \fft{\mu}{(\lambda^2+\mu^2)^{3/2}}\, \Omega_\2
\ ,\label{f5}\\
&&{\cal F}_\2 = dB=\Omega_\2 \ .\nn
\eea
%%%%

    We are now in a position to implement the T-duality transformation
from the type IIB description to the type IIA description in $D=5$.
 From appendix A, and (\ref{f5}), we see that the field strengths in the
type IIA picture will be
%%%%%
\bea
&& F_\3= \lambda\, \ep(\ads)\ ,\qquad
  \cF_\2 = \fft{\lambda}{(\lambda^2+\mu^2)^{3/2}}\, \Omega_\2\ ,\nn\\
&&F_{\3 1} = -\mu\, \ep(\ads)\ ,\qquad
F_\2 = \fft{\mu}{(\lambda^2+\mu^2)^{3/2}}\, \Omega_\2\ ,\label{ft5}\\
&&F_{\21} = \Omega_\2\ .\nn
\eea
%%%%%
 From (\ref{lambda}) and (\ref{varphi}), together with the fact that
we are taking $\phi_1=\phi_2=0$ in the original type IIB solution, it
follows that the dilatons in the type IIA picture will be given by
%%%%%
\be
e^{\varphi}= (\lambda^2+\mu^2)^{1/\sqrt6}\ ,\qquad
e^{\phi_1}=e^{\phi_2} = (\lambda^2+\mu^2)^{1/2}\ .
\ee

Finally, we can oxidise the type IIA solution that we have just
obtained back to $D=6$, by retracing the standard Kaluza-Klein
reduction steps.  Doing so, we find that the six-dimensional metric in
the type IIA picture is
%%%%%
\be
ds_6^2 = (\lambda^2+\mu^2)^{-1/2}\, ds^2(\ads) +
(\lambda^2+\mu^2)^{-3/2}\, \Big[ d\Omega_2^2 +
\fft{\lambda^2}{\lambda^2 + \mu^2}\, (dz' + B)^2 \Big]\ ,\label{2amet}
\ee
%%%%%
where $B$ is a potential such that $\Omega_\2 = dB$, and the
coordinate $z'$ is related to $z$ by
%%%%%
\be
z = \fft{\lambda}{(\lambda^2 +\mu^2)^{3/2}}\,
z' = Q_\ns\, z'\ .\label{zrel}
\ee
%%%%%
(The last equality follows from (\ref{charges}).)  It is
straightforward to verify that the area of the horizon of the metric
(\ref{2amet}) is the same as that before the Hopf T-duality
transformation, given by (\ref{area}).  The type IIA field strengths
in $D=6$ are given by
%%%%%
\bea
&&F_\4 = -\mu\, \ep(\ads)\wedge (dz+ \cA_\1) \ ,\qquad
F_\3 = \lambda\, \ep(\ads) + \Omega_\2\wedge (dz+\cA_\1)\ ,\nn\\
&&F_\2 = \fft{\mu}{(\lambda^2+\mu^2)^{3/2}}\, \Omega_\2\ ,\label{2afield}
\eea
%%%%%
where
%%%%%
\be
{\cal A}_\1 = \fft{\lambda}{(\lambda^2+\mu^2)^{3/2}}\, B =
Q_\ns\, B\ .\label{bredef}
\ee
Note that in the above T-duality transformation the Buscher rules
\cite{buscher} are insufficient, since we have R-R fields as well as
NS-NS fields involved in the solution.  For this reason, we have
constructed the two low-energy effective actions and explicitly
derived the T-duality transformations that relate them.

     We find that the charges carried by these field strengths are as
follows:
%%%%%
\bea
Q^\3_{\rm elec} &\equiv & \ft1{16\pi^2}\int_{S^3}\, e^{-\phi_1-\phi_2}\,
{*F_\3} = Q_\ns\ ,\nn\\
Q^\3_{\rm mag} &\equiv & \ft1{16\pi^2}\int_{S^3}\,
F_\3 = 1\ ,\nn\\
Q^\4_{\rm elec} &\equiv & \ft1{4\pi}\int_{S^2}\, e^{\fft12\phi_1-
\fft32\phi_2}\, {*F_\4} = - Q_\rr\ ,\nn\\
Q^\2_{\rm mag} &\equiv & \ft1{4\pi}\int_{S^2} \,
F_\2 = Q_\rr\ .\label{2acharges}
\eea
%%%%%
If the fibre coordinate $z'$ in (\ref{2amet}) had had the period
$4\pi$, then the topology of the compact 3-space would have been
$S^3$.  Since it is related to $z$ as given in (\ref{zrel}), and $z$
has period $4\pi$, it follows that $z'$ has period $4\pi/Q_\ns$, and
hence the topology of the compact 3-space is $S^3/Z_{Q_\ns}$, the
cyclic lens space of order $Q_\ns$.

On the other hand the magnetic charge
carried by the field strength $F_\3$ is equal to 1, having started, in
the original solution, as $Q_\ns$.  This is because the T-duality has
exchanged the original charge $Q_\ns$ with the unit charge
corresponding to the unit topological charge of the $U(1)$ bundle over
$S^2$ that describes the 3-sphere.
     Furthermore, we can see from (\ref{2amet}) that the metric on the
lens space is not in general the ``round'' one, but is instead
squashed along the $U(1)$ fibre direction, with a squashing factor
$\nu$ given by
%%%%%
\be
\nu = \fft{\lambda}{\sqrt{\lambda^2 +\mu^2}} =
\fft{Q_\ns}{\sqrt{Q_\ns^2 + Q_\rr^2}}\ .
\ee
In other words, the metric on the compact 3-space is proportional to
\be
ds^2 = \ft14 d\Omega_2^2 + \ft14 \nu^2\, (dz+B)^2\ ,
\ee
whose Ricci tensor, in the natural orthonormal basis, is given by
\be
R_{ab} = {\rm diag}\, (4-2\nu^2, 4-2\nu^2, 2\nu^2)\ .
\ee
When $\nu=1$, the metric is Einstein and the 3-sphere or lens space is
``round.''

As we mentioned earlier, we could have considered original solutions
in which the constant dilatons $\phi_1$ and $\phi_2$ were non-zero, in
which case the original electric and magnetic charges need not have
been equal.  The lens space after the Hopf T-duality transformation
will then be $S^3/Z_{Q_\ns^{\rm mag}}$.  Also, we can generalise the
starting point further by consider a solution on the product of
AdS$_3$ and the lens space $S^3/Z_n$, rather than simply AdS$_3\times
S^3$.  (From the lower-dimensional point of view, this corresponds to
giving the Kaluza-Klein vector a magnetic charge $n$ rather than 1.)
If we do this, then we find that a solution AdS$_3\times S^3/Z_n$ for
(\ref{2blag}), with charges $Q_\ns^{\rm elec}$, $Q_\ns^{\rm mag}$,
$Q_\rr^{\rm elec}$ and $Q_\rr^{\rm mag}$ will result, after the
T-duality transformation, in a solution AdS$_3\times S^3/Z_{Q_{\rm
    mag}^\ns}$ for (\ref{2alag}) with charges
%%%%%
\be
Q^\3_{\rm elec}= Q_\ns^{\rm elec}\ ,\qquad
Q^\3_{\rm mag}= n\ ,\qquad
Q^\4_{\rm elec}= - Q_\rr^{\rm elec}\ ,\qquad
Q^\2_{\rm mag}= Q_\rr^{\rm mag}\ .\label{2acharges2}
\ee

To summarise, we see that if we consider the case where the original
AdS$_3\times S^3/Z_n$ solution in the type IIB description carries
only an NS-NS charge whose magnetic component is $Q_{\rm mag}^\ns=m$,
and so $Q_\rr=0$, then after the Hopf T-duality transformation the
solution for (\ref{2alag}) will be of the form AdS$_3\times S^3/Z_m$,
where the metric on the cyclic lens space is still the ``round'' one,
and the magnetic charge becomes $Q_{\rm mag}^\ns = n$.  In the special
case where $n=m$ and $Q_\rr=0$, the AdS$_3\times S^3/Z_n$ solution is
invariant under the Hopf T-duality.  If, on the other hand, we start
with an AdS$_3\times S^3/Z_n$ solution with only R-R charges, then
after the Hopf T-duality transformation the $S^3/Z_n$ is completely
``untwisted,'' and the solution will be of the form AdS$_3\times
S^1\times S^2$.  (This is analogous to the untwisting of AdS$_5\times
S^5$ discussed in \cite{DLP2} and the untwisting of AdS$_4 \times
S^7$ discussed in \cite{swos}.  The untwisting of $S^3$ to $S^2\times
S^1$ was also discussed in \cite{lag}.)  In the generic case where the
original AdS$_3\times S^3/Z_n$ solution carries both NS-NS and R-R
charges, then after the T-duality transformation the solution will be
of the form AdS$_3\times S^3/Z_m$, where the metric on the compact
3-space space is not only factored now by $Z_m$, but it is also
squashed.

Although the construction of conformal field theories with background
R-R charges is problematical, there is an exact CFT duality 
statement\footnote{We are grateful to Costas Bachas for urging us to 
provide a CFT proof.  A CFT discussion of the $D=10$ superstring 
compactification 
on $S^{3} \times S^{3}$ down to AdS$_{2} \times S^{2}$ may be found 
in \cite{antoniadis}} 
in the case of pure NS-NS charge, \ie when $\mu=0$.  This can be seen
by looking at the original solution (\ref{adssol}), and the final
Hopf-dualised solution (\ref{2amet}, \ref{2afield}), in the string-frame metric
$ds_6^2({\rm string})$. This is related to the six-dimensional
Einstein-frame metric $ds_6^2$ by $ds_6^2({\rm string})=
e^{\fft12(\phi_1+\phi_2)}\, ds_6^2$.  Thus we find that the original
solution can be written as
\bea
ds_6^2({\rm string}) &=& ds^2(AdS) + \lambda^{-2}\, \Big[ d\Omega_2^2
+ (dz+B)^2\Big] \ ,\nn\\
F_\3 &=& \lambda^{-2}\, \Sigma_\3 + \lambda^{-2}\, \Omega_2\wedge
(dz+B)\ ,
\eea
where $\Sigma_\3$ is the volume form of the ``unit'' AdS$_3$, and that
correspondingly in the final Hopf-dualised solution we have
\bea
ds_6^2({\rm string}) &=& ds^2(AdS) + \lambda^{-2}\, \Big[ d\Omega_2^2
+ (dz'+B)^2\Big] \ ,\nn\\
F_\3 &=& \lambda^{-2}\, \Sigma_\3 + \lambda^{-2}\, \Omega_2\wedge
(dz'+B)\ ,
\eea
where $z'=\lambda^2\, z =Q_\ns^{-1}\, z$.  The T-duality can be
understood in this case from the standard Buscher rules \cite{buscher} 
applied to the $SU(2)$ WZW model.
In general, one has the statement that a solution on $S^3/Z_m$ with
3-form flux $n$ is Hopf dual to a solution on $S^3/Z_n$ with
3-form flux $m$.  This is because the Kaluza-Klein vector in the
$S^3/Z_m$ solution carries $m$ units of charge, whereas the winding 
mode vector coming from the 3-form carries $n$ units of charge.  
The invariance under T-duality in the special case $n=m=1$ is discussed in 
\cite{kiritsis}.

\section{AdS$_3$ (un)twisted and squashed}

     In the same way as odd-dimensional spheres can be viewed as
$U(1)$ bundles over complex projective spaces, so we can view
odd-dimensional AdS spacetimes as bundles over certain
Lorentzianisations of the complex projective spaces.  The case of
AdS$_3$ is particularly simple, since in this case the base space is
nothing but AdS$_2$.

Let us begin by considering the unit $S^3$, written as a $U(1)$ bundle
over $S^2$:
%%%%%
\be
ds^2 = \ft14 d\theta^2 + \ft14\sin^2 \theta\, d\phi^2 + \ft14
(d\psi+\cos\theta d\phi)^2\ .\label{s3met}
\ee
%%%%%
Now perform an analytic continuation to a Lorentzian signature, by
sending:
%%%%%
\be
\theta\longrightarrow \ft12 \pi - i\, \rho\ ,\qquad
\psi\longrightarrow i\, x\ ,\qquad \phi\longrightarrow t\ .
\ee
%%%%%
This gives us the metric (after making an overall sign change to go from
east-coast to west-coast notation)
%%%%%
\be
ds^2 = -\ft14 \cosh^2 \rho \, dt^2 + \ft14 d\rho^2 + \ft14 (dx + \sinh\rho\,
dt)^2\ .\label{ads}
\ee
%%%%%
It is not hard to calculate the curvature for this metric, and to
verify that it has Ricci tensor $R_{\mu\nu} = -2\, g_{\mu\nu}$.  Thus
it is AdS$_3$, since the cosmological constant is negative.  Note that
$t$ is periodic, $0\le t\le 2\pi$, but $\rho$ and $x$ both range over
the entire real line.  The parameterisation of points in AdS$_3$,
viewed as the $O(2,2)$-invariant hyperboloid $X_1^2+ X_2^2 - X_3^2
-X_4^2 =1$ in $\R^4$, can be given in terms of $t$, $\rho$ and $x$ as
follows:
%%%%%
\bea
\pmatrix{X_1\cr X_2} &=& \ft1{\sqrt2}
\pmatrix{\cos \ft12 t & \sin \ft12 t\cr
          -\sin \ft12 t & \cos\ft12 t} \, \pmatrix{\cosh x_-\cr \cosh
          x_+}\ ,\nn\\
&&\nn\\
\pmatrix{X_3\cr X_4} &=& \ft1{\sqrt2}
\pmatrix{\cos \ft12 t & \sin \ft12 t\cr
          -\sin \ft12 t & \cos\ft12 t} \, \pmatrix{\sinh x_-\cr \sinh
          x_+}\ ,
\eea
%%%%%
where $x_\pm \equiv\ft12 (x\pm \rho)$.  It can be shown that the
coordinates $t$, $\rho$ and $x$ give a 1-1 mapping to points on the
hyperboloid.

    It is evident that we can view (\ref{ads}) as a bundle over
a base space which is AdS$_2$, for which the unit-radius metric is
$d\Sigma_2^2 = -\cosh^2\rho\, dt^2 + d\rho^2$.  (We shall in
general use $d\Sigma_n^2$ to denote the ``unit'' AdS$_n$ metric,
which is defined to be the one whose Ricci tensor has the form
$R_{\mu\nu} = -(n-1)\, g_{\mu\nu}$.  This is analogous to the
definition of the unit $n$-sphere.)  Thus we have
%%%%%
\be
d\Sigma_3^2 = \ft14 d\Sigma_2^2 + \ft14 (dx+\tB)^2\ ,
\ee
%%%%%
where $d\tB = \Sigma_\2$.  (We are defining $\Sigma_\n$ to be the
volume form on the unit AdS$_n$.)

    Let us note here that there is also another closely related
metric on AdS$_3$, which also allows one to do a reduction on
the fibre coordinate $x$, namely
%%%%%
\be
ds^2 = -\ft14 e^{2\rho}\, dt^2 + \ft14 d\rho^2 + \ft14
(dx+ e^{\rho} \,dt)^2\ .\label{ads2}
\ee
%%%%%
This can be shown to be an AdS$_3$ metric with $R_{\mu\nu} = -2\,
g_{\mu\nu}$.  It is interesting because it arises by taking the
near-horizon limit of the boosted dyonic string in $D=6$ (see appendix
C).  (In other words, the intersection of the dyonic string with a
wave.  If reduced to $D=5$, this boosted dyonic string becomes a
3-charge black hole; {\it i.e.}\ the Reissner-Nordstr{\o}m black
hole.)

    The coordinate transformation that relates the two AdS$_3$ metrics
(\ref{ads}) and (\ref{ads2}) is the following:
%%%%%
\bea
e^{\td \rho} &=& \sinh\rho + \cosh\rho\, \cos t\ ,\nn\\
\td t\, e^{\td \rho} &=& \cosh\rho\, \sin t\ ,\label{coordtrans}\\
e^{\td x} &=&
\fft{(e^\rho\, \cot \ft12 t -1)\, e^x}{(e^\rho\, \cot \ft12 t +1)}\ ,\nn
\eea
%%%%%
where the tilded coordinates here denote the coordinates in the metric
(\ref{ads2}).  An important feature
of both (\ref{ads}) and (\ref{ads2}) is that there is no
coordinate-dependent function multiplying the vielbein in the fibre
direction, and so the circle in the $U(1)$ reduction will have a
constant length.

This Hopf T-duality on the fibres of AdS$_3$ should be contrasted with
the T-duality discussed in \cite{horowitz,andreas}, or with a
T-duality on one of the horospherical coordinates of AdS.  (See also 
\cite{nk1}.)  In these
two cases the metric is already diagonal, and the size of the
compactifying circle is not constant, but instead depends on other
coordinates of the AdS.  It follows that after T-dualisation, the dual
theory has a dilaton that is singular on the horizon, and hence so
also is the metric.  By contrast, in the Hopf dualisation of AdS$_3$
discussed above, the constant radius of the circle implies that the
dilaton is non-singular, and the metric has no local curvature
singularity.  The difference is further highlighted by the analysis of
the supersymmetry.  In the case of the horospherical T-duality or the
T-duality in \cite{horowitz,andreas}, supersymmetry is always (at
least partially) broken.  In the case of Hopf T-duality on AdS$_3$,
however, the supersymmetry is either all preserved or all broken,
depending on an orientation choice in the Hopf fibering.  We shall
discuss this, and make detailed comparisons between the various
T-dualities, in section 6.

Note that we can also have the concept of ``squashing'' for AdS$_3$
where the length of the fibres
is rescaled relative to the size of the base AdS$_2$.  As in the case
of the sphere, this will be an homogeneous squashing.  Thus we may
consider a squashed AdS$_3$ metric
%%%%%
\be
ds^2 = \ft14 d\Sigma_2^2 + \ft14\, \nu^2\, (dx+\tB)^2\ ,
\ee
%%%%%
where $\nu$ is a constant squashing parameter.  The vielbein
components of the Ricci tensor in the natural orthonormal basis are
%%%%%
\be
R_{ab} =\hbox{diag}\, (4- 2\nu^2, 2\nu^2 -4, - 2\nu^2)\ ,
\ee
%%%%%
where the first entry is the $R_{00}$ component.  More generally,
there also exist squashed versions of AdS$_{2n+1}$ for any $n$, of the
form $ds^2({\rm AdS}_{2n+1}) = ds^2(\wtd{CP}^n) + \nu^2\, (dx +
\tB)^2$, where $\wtd{CP}^n$ denotes a Lorentianisation of $CP^n$, and
$d\tB$ is the volume form on $\wtd{CP}^n$.

All the previous steps of dimensional reduction on the fibres, which
we did for the $S^3$ factor in AdS$_3\times S^3$, can now be repeated
for the AdS$_3$ itself.  The computations are essentially identical,
since the dimension of the AdS$_3$ is the same as that of the $S^3$,
with the exception of the details of the field strengths.  The final
result is that the solution AdS$_3\times S^3$ given by (\ref{adssol})
becomes, after performing a T-duality transformation on the AdS$_3$
fibre coordinate $x$,
%%%%%
\be
ds_6^2 =
(\lambda^2+\mu^2)^{-3/2}\, \Big[ d\Sigma_2^2 +
\fft{\lambda^2}{\lambda^2 + \mu^2}\, (dx' + \tB)^2 \Big]
+(\lambda^2+\mu^2)^{-1/2}\, ds^2(S^3)
\ ,\label{2aadsmet}
\ee
%%%%%
where $\tB$ is a potential such that $\Sigma_\2 = d\tB$, and the
coordinate $x'$ is related to $x$ by
%%%%%
\be
x = \fft{\lambda}{(\lambda^2 +\mu^2)^{3/2}}
\, x' = Q_\ns\, z'\ .\label{xrel}
\ee
%%%%%
The type IIA field strengths in $D=6$ are given by
%%%%%
\bea
&&F_\4 = -\mu\, Q_\ns \, \ep(S^3)\wedge (dx' + \tB) \ ,\qquad
F_\3 = \lambda\, \ep(S^3) + Q_\ns\, \Sigma_\2\wedge (dx'+\tB)\ ,\nn\\
&&F_\2 = Q_\rr\, \Sigma_\2\ .
\eea
%%%%%
In particular, we see that in the case where the solution is supported
purely by R-R 3-form charges, then after doing the T-duality
transformation we arrive at an ``untwisted'' solution AdS$_2\times S^1
\times S^3$.  If the solution instead carries only NS-NS charges, then
the structure of the T-duality transformed solution is essentially
unchanged, in the sense that there are just overall rescalings
$1/\lambda$ and $1/\lambda^3$ of the 3-sphere and AdS$_3$ factors in
the metric.  Since the fibre coordinate $x$ here ranges over the
entire real line, we do not really have the notion of a ``lens space''
for AdS.  We could consider instead the spacetime that one
obtains by taking the fibre coordinate $x$ to be periodic.  This will
no longer be globally AdS$_3$, but instead only a part of it.
If we define a spacetime $U_n$ to be AdS$_3$ with $x$ having a period
$L/n$, for some specified $L$, then one could say that the T-dual of
the solution $U_1\times S^3$ carrying purely NS-NS charge $Q_\ns$ is
the solution $U_{Q_\ns}\times S^3$.  As in the case of the Hopf
T-duality for $S^3$, all of the above discussion generalises
straightforwardly to the case where we allow the dilaton moduli to be
non-zero, so that there can be independent electric and magnetic
charges for each field strength.

    Note that the area of the horizon is preserved under the Hopf
T-duality on the fibre coordinate $x$ of the AdS$_3$.  It is given by
$A\sim (\lambda^2+\mu^2)^{-2}$.

\section{AdS$_3$ and $S^3$ (un)twisted and squashed}

     We may now put together the results of the previous two sections,
by starting with an AdS$_3\times S^3$ solution (\ref{adssol}),
performing a T-duality transformation first using the $U(1)$ fibres of
$S^3$, and then performing a second T-duality transformation using the
fibres of AdS$_3$.  After doing this, we find that the final metric is
given by
%%%%%
\bea
ds_6^2 =\fft{1}{(\lambda^2+\mu^2)^2}\, \Big[ d\Sigma_2^2 +
\fft{\lambda^2}{\lambda^2+\mu^2} \, (dx'+\tB)^2 + d\Omega_2^2 +
\fft{\lambda^2}{\lambda^2 +\mu^2}\, (dz' + B)^2 \Big]\ ,
\eea
%%%%%
where $z'$ is given by (\ref{zrel}) and $x'$ is given by (\ref{xrel}).
The field strengths in the final solution are given by
\bea
F_\3^\ns &=& \Omega_\2 \wedge (dz+{\cal A}_\1) + \Sigma_\2\wedge
(dx+\wtd{\cal A}_\1)\ ,\nn\\
F_\3^\rr &=& Q_\rr\, \Omega_\2 \wedge (dx+\wtd{\cal A}_\1)
    -Q_\rr\, \Sigma_\2 \wedge (dz+ {\cal A}_\1)\ ,
\eea
%%%%%
where ${\cal A}_\1$ is given by (\ref{bredef}) and
%%%%%
\be
\wtd{\cal A}_\1 = \fft{\lambda}{(\lambda^2+\mu^2)^{3/2}}\, \tB =
Q_\ns\, \tB\ .
\ee
%%%%%

      If the NS-NS charge is zero, then in this doubly-transformed
solution both the $S^3$ and the AdS$_3$ are untwisted, giving
(AdS$_2\times S^1)\times (S^2\times S^1)$.  If instead the R-R charge
is zero, the solution becomes twisted to $U_{Q_\ns} \times
(S^3/Z_{Q_\ns})$.  When the NS-NS and R-R charges are both present,
the solution is squashed and twisted in both the AdS$_3$ and
$S^3$ factors.  Note that the squashing parameter is the same for both
the AdS$_3$ and the $S^3$.

Again in this case, the Hopf T-duality transformation preserves the
area of horizon, given by $A\sim (\lambda^2 +\mu^2)^{-2}$.  It follows
that the black hole entropy, which is a quarter of the area, is also
invariant under the Hopf T-duality.  In the case of a 5-dimensional
black hole that oxidises to a boosted dyonic NS-NS string in $D=6$,
this invariance can be easily understood:  It follows from
(\ref{nh62}) in this case that the black hole entropy is given
by $S\sim \sqrt{Q_w\, q\, p}= \sqrt{n\, Q_e^{\ns}\, Q_m^\ns}$.
As we have seen earlier, Hopf T-duality has the effect of
interchanging $n$ and $Q_e^{\ns}$, and hence the entropy is left invariant.
In this paper, we have shown that in general the entropy of the
black hole is preserved under Hopf T-duality, even when it is
supported by R-R as well as NS-NS charges.  Thus even though the
metric on the AdS$_3$ or $S^3$ may be (un)twisted and squashed by the
transformation, the area of the horizon is preserved.

\section{Supersymmetry and Killing spinors}

It is well known that T-duality transformations can break
supersymmetries of $p$-brane solutions, at the level of the low-energy
effective supergravity.  For example, the near-horizon limit of the
D3-brane in ten dimensions, which is of the form AdS$_5\times S^5$ and
hence preserves all the supersymmetry, is T-dual to the near-horizon
limit of a D4-brane, which is a product of a domain wall and a sphere
(rather than an AdS and a sphere), and which breaks half of the
supersymmetry.  The phenomenon of supersymmetry breaking, at the level
of supergravity, has also been seen in the Hopf T-duality
transformations discussed in \cite{DLP2}.  For example, the Hopf
dualisation of AdS$_5\times S^5$ to AdS$_5\times CP^2\times S^1$
breaks all the supersymmetry at the supergravity level \cite{DLP2}.
In another example, the Hopf reduction of the AdS$_4\times S^7$
solution of $D=11$ supergravity to the AdS$_4\times CP^3$ of $D=10$
type IIA supergravity breaks either all of the eight supersymmetries,
or two out of the eight, depending on the orientation of the internal
manifold \cite{np,swos}.

Of course all these statements about supersymmetry breaking are made
at the level of the massless Kaluza-Klein modes in the supergravity
theory.  In some cases, the supersymmetry will be restored when one
includes the Kaluza-Klein massive modes or the string winding modes.
To analyse the behaviour of supersymmetry under T-duality, one should
separate the discussion into two parts: firstly, compactification on a
circle, and secondly, the T-duality transformation itself.
Supersymmetry breaking, if it occurs at all, is a consequence of the
compactification on the circle.  The T-duality transformation in the
full string theory always preserves whatever supersymmetry has
survived the circle compactification.  An example where
compactification breaks supersymmetry is an AdS spacetime, written in
horospherical coordinates $ds^2 = d\rho^2 + e^{2\rho}\,
\eta_{\mu\nu}\, dx^\mu\, dx^\nu$, and compactified on one of the
spatial $x^\mu$ coordinates.  Although there is a translational
isometry, thus allowing a circle compactification, one finds that half
of the Killing spinors on the AdS spacetime depend linearly on
$x^\mu$, and thus the supersymmetries associated with these Killing
spinors will be broken once the chosen $x^\mu$ coordinate is taken to
be periodic \cite{lpt}.  The other half of the Killing spinors are
independent of $x^\mu$, and so half of the supersymmetries survive the
compactification.  This statement is true even after including all the
massive Kaluza-Klein modes.  A T-duality transformation will not
result in any further breaking (or restoring) of supersymmetry.

   A contrasting example is provided by the Hopf reduction on the
$U(1)$ fibres of $S^5$ in the AdS$_5\times S^5$ solution of the type
IIB theory.  The fibre coordinate here is naturally periodic.
Although at the level of the massless Kaluza-Klein modes the
compactification on the $U(1)$ fibres breaks all the supersymmetry, it
is restored once the massive Kaluza-Klein modes are included.  If we
perform a T-duality transformation on the fibre coordinate, the
AdS$_5\times S^5$ solution is mapped to an AdS$_5\times CP^2\times S^1$
solution in the type IIA theory.  At the level of supergravity, even
when the massive Kaluza-Klein modes are included, this type IIA
solution breaks all the supersymmetries.  However in the full string
theory, where the string winding modes are also included, the
full supersymmetry is reinstated \cite{DLP2}.

Thus, for example, we see that there are two different kinds of
T-duality transformations that can be performed on a D3-brane.  One of
these maps the D3-brane to a D4-brane.  In this case, the near-horizon
AdS$_5\times S^5$ limit of the D3-brane is mapped to a product of a
domain wall and a 4-sphere, which is the near-horizon limit of the
D4-brane.  In this case it is one of the horospherical coordinates
$x^\mu$ in AdS$_5$ that is periodically identified and used in the
T-duality transformation.  (Since the $x^\mu$ coordinates are in fact
the world-volume coordinates of the D3-brane.)  Thus half of the
supersymmetry is already broken in the process of making the
identification.  In other words, the apparent discrepancy between the
``enhanced supersymmetry'' seen in the near horizon limit of an
unwrapped D3-brane and the usual $1/2$ supersymmetry of the
near-horizon limit of the D4-brane is not the result of any breaking
of supersymmetry by T-duality.  Rather, it is simply a consequence of
the fact that there {\it is} no ``supersymmetry enhancement'' in the
near-horizon limit of a wrapped D3-brane \cite{lpt}.
The other kind of T-duality that can be performed on a D3-brane is on
the fibre coordinate of the foliating 5-spheres in its 6-dimensional
transverse space.  In this case, the fibre coordinate is naturally
periodic, and the ``supersymmetry breaking'' is a pure artifact of the
low-energy supergravity approximation.  In the full string theory,
when the Kaluza-Klein and winding modes are included, there is no
supersymmetry breaking at all \cite{DLP2}.

    At the level of the massless modes in the reduction of the
supergravity theory, the Hopf T-duality transformations on
AdS$_3\times S^3$ that we have been considering in this paper either
preserve all the supersymmetry, or they break all the supersymmetry,
depending upon an orientation associated with the Hopf reduction.  In
this section, we shall demonstrate this by giving an explicit
construction of the Killing spinors on AdS$_3$ and $S^3$.  These have
been given previously \cite{lpt,fy,lpr}, but in coordinate systems
that are not convenient for our present purposes. The relation between
the coordinates used in \cite{lpr}, in which the metric on the unit
3-sphere is $d\Omega_3^2 = d\theta_3^2 + \sin^2\theta_3\, (d\theta_2^2
+ \sin^2\theta_2\, d\theta_1^2)$, and the coordinates in
(\ref{s3met}), is $\theta_1=\ft12(\psi+\phi)$, $\cot\theta_2 =
\tan\ft12\theta\, \cos\ft12(\psi-\phi)$, $\cos\theta_3 =
\sin\ft12\theta\, \sin\ft12(\psi-\phi)$.  In principle, the Killing
spinors obtained in \cite{lpr} can be re-expressed in terms of the
coordinates of the metric (\ref{s3met}) using these relations, but the
result will be in an inconvenient local Lorentz frame.

    We begin by constructing the Killing spinors for the unit $S^3$,
using the metric given in (\ref{s3met}).  The vielbein and spin
connection are given by
%%%%%
\bea
&&e^1 = \ft12 d\theta\ ,\qquad
e^2= \ft12 \sin\theta\, d\phi\ ,\qquad
e^3 = \ft12(d\psi + \cos\theta\, d\phi)\ ,\nn\\
&&\omega_{23} = -e^1\ ,\qquad
\omega_{31} = -e^2\ ,\qquad
\omega_{12}= -2\cot\theta\, e^2 + e^3\ .\label{vom}
\eea
%%%%%
The Killing spinor equation is $D_\mu\, \ep^\pm = \pm\ft{\im}{2}\,
\Gamma_\mu\, \ep^\pm$.  In a given choice of conventions in the
AdS$_3\times S^3$ supergravity solution, either the $\ep^+$ Killing
spinors or the $\ep^-$ Killing spinors will be the ones that are
associated with unbroken supersymmetries.  Since the round $S^3$ is
invariant under orientation reversal, equal numbers of Killing spinors
$\ep^+$ and $\ep^-$ exist (namely two of each).  Substituting
(\ref{vom}) into the Killing spinor equation, we find, in the basis
where $\Gamma_1=\sigma_3$, $\Gamma_2=\sigma_1$ and
$\Gamma_3=\sigma_2$, with $\sigma_i$ denoting the standard Pauli
matrices, that the two sets of Killing spinors are
%%%%%
\bea
&&\ep^+_1= \pmatrix{e^{\ft{\im}{2}(\phi+\theta)} \cr
                 \im\,  e^{\ft{\im}{2}(\phi-\theta)} }\ ,\qquad
\ep^+_2= \pmatrix{e^{\ft{\im}{2}(\theta-\phi)} \cr
                 -\im\,  e^{-\ft{\im}{2}(\theta+\phi)} }\ ,\label{epp}\\
&&\nn\\
&&\ep^-_1 = \pmatrix{e^{\ft{\im}{2}\psi} \cr
                     -\im\, e^{\ft{\im}{2}\psi} }\ ,\qquad
\ep^-_2 = \pmatrix{e^{-\ft{\im}{2}\psi} \cr
                     \im\, e^{-\ft{\im}{2}\psi} }\ ,\label{epm}
\eea
%%%%%
Thus we see that if the conventions have been chosen so that the
$\ep^+$ Killing spinors are associated with unbroken supersymmetries,
then they survive the dimensional reduction to $D=5$ since they are
independent of $\psi$.  Under these circumstances, the supergravity
solution after the Hopf T-duality transformation on the $\psi$
coordinate will still be fully supersymmetric.  If, on the other hand,
the conventions have been chosen so that the $\ep^-$ Killing spinors
are associated with unbroken supersymmetries, then they will not
survive the dimensional reduction process since they both depend on
$\psi$.  In this case, all the supersymmetry will be broken in the
supergravity solution after the Hopf T-duality transformation.  Note
that in our discussion here, we have made a fixed orientation choice
for our Hopf reduction (implicit in our definition of the vielbeins in
(\ref{vom}), and so the two supersymmetry possibilities arise from two
possible convention choices for the original supergravity solution.
Equivalently, one could think of making a fixed convention choice in
the original solution, and the two supersymmetry possibilities would
then arise from the two possible orientations in the Hopf reduction.

In order to discuss the supersymmetry in the case of the T-duality
transformations on the Hopf fibres of AdS$_3$, we need first to give
an analogous construction of the Killing spinors in AdS$_3$.  On the
unit AdS$_3$, these satisfy $D_\mu\, \ep^\pm = \pm\ft12 \Gamma_\mu\,
\ep^\pm$.  Let us first consider the metric (\ref{ads}), for which the
vielbein and spin connection will be
%%%%%
\bea
&&e^0 = \ft12 \cosh\rho\, dt\ ,\qquad
e^1= \ft12 d\rho \ ,\qquad
e^2 = \ft12(dx + \sinh\rho\, dt)\ ,\nn\\
&&\omega_{01}= -2\tanh\rho\, e^0 + e^2\ ,\qquad
\omega_{02} = e^1\ ,\qquad
\omega_{12} = -e^0\ .\label{vomads}
\eea
%%%%%
Taking the Dirac matrices to be $\Gamma_0= -\im\, \sigma_1$,
$\Gamma_1=\sigma_3$ and $\Gamma_2=\sigma_2$, we find that the
solutions for the Killing spinors are
%%%%%
\bea
&&\ep^+_1 = \pmatrix{e^{\ft12(\rho+\im t)} \cr
                     -e^{\ft12(-\rho+\im t)}}\ ,\qquad
\ep^+_2 = \pmatrix{e^{\ft12(\rho-\im t)} \cr
                     e^{-\ft12(\rho+\im t)}}\ ,\label{eppads}\\
&&\nn\\
&&\ep^-_1 = \pmatrix{e^{\ft12 x} \cr
                    -\im\, e^{\ft12x}}\ ,\qquad
\ep^-_2 = \pmatrix{ e^{-\ft12 x} \cr
                    \im\, e^{-\ft12 x} }\ .\label{epmads}
\eea
%%%%%
Again we see that the Hopf T-duality transformation on the fibre
coordinate $x$ will preserve either all or none of the supersymmetry,
depending upon the orientation.

    It is also instructive to construct the Killing spinors for the
AdS$_3$ metric (\ref{ads2}), since this is the one that arises as the
near-horizon limit of the boosted dyonic string.  For this, the
vielbein and spin connection will be:
%%%%%
\bea
&&e^0=\ft12 e^\rho\, dt\ ,\qquad e^1 = \ft12 d\rho\ ,\qquad
 e^2= \ft12(dx + e^\rho\, dt)\ ,\nn\\
&&\omega_{01} = -2\, e^0 + e^2\ ,\qquad \omega_{02} = e^1\ ,\qquad
\omega_{12} = - e^0\ .
\eea
%%%%%
Elementary calculations then show that the Killing spinors
$\ep^\pm$, satisfying $D_\mu\, \ep^\pm = \pm\ft12\, \Gamma_\mu\,
\ep^\pm$, are given by
%%%%%
\bea
&&\ep^+_1 = \pmatrix{t\, e^{\fft12\rho} \cr \im\, e^{-\fft12\rho}} \ ,
\qquad
\ep^+_2 = \pmatrix{e^{\fft12\rho} \cr 0}\ ,\\
&&\nn\\
&& \ep^-_1 = \pmatrix{e^{\fft12 x} \cr -\im\, e^{\fft12 x}}\ ,\qquad
\ep^-_2 =\pmatrix{e^{-\fft12 x} \cr \im\, e^{-\fft12 x}}\ .\label{epmads2}
\eea
%%%%%
Here also, we see that the two $\ep^+$ Killing spinors are independent
of the fibre coordinate $x$, while the two $\ep^-$ Killing spinors
depend on $x$.  It is worth remarking that the metric (\ref{ads2}) is
in some ways reminiscent of the horospherical metric on AdS$_3$, and
indeed this reflects itself in certain similarities in the form of
the solutions for the Killing spinors \cite{lpt}.

The orientation dependence of the supersymmetry is echoed in $D=4$,
where 4-charge black hole solutions, can either preserve $1/8$ of the
supersymmetry or break it entirely, depending on the relative signs of
the charges \cite{lpmult,khuri}.  These black holes are all related by
U-duality to a black hole in which two of the charges, one electric and one
magnetic, are carried by the two Kaluza-Klein vectors coming from the
reduction from $D=6$.  Upon oxidation to $D=6$, this gives a metric
whose near-horizon limit is AdS$_3\times S^3$, with the two
Kaluza-Klein vectors providing the twisting of the $S^3$ and the
AdS$_3$ fibres.  Thus the sign of the fibre orientations is precisely
related to the signs of the charges in $D=4$.

The above discussion is at the level of the massless Kaluza-Klein
modes in the supergravity theory.  If the orientation for the Hopf
fibration is such that the relevant Killing spinors do not depend on
the fibre coordinate, then statement of preservation of full
supersymmetry under the Hopf T-duality extends to the full string
theory.  If the orientation is opposite, so that all the relevant
Killing spinors depend on the fibre coordinate, then the discussion
bifurcates into two categories in the full string theory.  If the Hopf
reduction is on the (naturally periodic) $U(1)$ fibres of the $S^3$,
then, as we see from (\ref{epm}), the Killing spinors will be included
once the non-zero modes in the Kaluza-Klein Fourier expansions are
taken into account.  In this case, the full supersymmetry preservation
under Hopf T-duality is reinstated by including all the Kaluza-Klein
modes.  On the other hand, if the Hopf reduction is on the fibre
coordinate of AdS$_3$, then, as we can see from (\ref{epmads}) or
(\ref{epmads2}), the Killing spinors will still be excluded from the
spectrum even after the Kaluza-Klein non-zero modes are included.  In
this case, therefore, the supersymmetry remains broken even in the
full string theory.

\section{Non-dilatonic black holes, and Hopf T-duality}

     Non-dilatonic black holes (for which the dilatons are finite
on the horizon) in maximal supergravity arise in $D=5$ and $D=4$.  In
$D=5$, they are supported by three field strengths, and associated
with each is an independent harmonic function.  There are a total of
45 possible field configurations that can give rise to
five-dimensional non-dilatonic black holes \cite{classp}, {\it viz.}
%%%%%%%%%
\be
D=5:\quad \{ F_{\2 ij}, F_{\2 k\ell}, F_{\2 mn} \}_{15}\ ,\quad
       \{ *F_{\3 i}, {\cal F}_\2^j, F_{\2 ij} \}_{30}\ .
\label{d5n3}
\ee
%%%%%%%
(We are using the notation of \cite{lpsol,cjlp1,classp} here.  The
subscripts on each set of field strengths denotes the multiplicities
of the solutions.  The Hodge duals indicate that the associated fields
carry electric charges if the fields without duals carry magnetic
charges, and {\it vice versa}.)  In $D=4$, non-dilatonic black holes
are supported by four field strengths, and can arise with the
following possible field-strength configurations \cite{classp}:
%%%%%%%
\bea
N=4:&& \{ F_{\2 ij}, F_{\2 k\ell}, F_{\2 mn}, *{\cal F}_\2^p
        \}_{105+105}\ ,\quad
       \{F_{\2 ij}, *F_{\2 ik}, {\cal F}_\2^j, *{\cal
         F}_\2^k \}_{210}\ ,\nonumber\\
    && \{ F_{\2 ij}, F_{\2 k\ell}, *F_{\2 ik}, *F_{\2 j\ell}
        \}_{210}\ .\label{d4n4}
\eea
%%%%%%%
The near-horizon regions of these $D=5$ and $D=4$ black holes are
AdS$_2\times S^3$ and AdS$_2\times S^2$ respectively.  If we
dimensionally oxidise these solutions to $D=6$, they will describe the
intersections of $p$-branes, waves and NUTS.  There are four
possible near-horizon limits that can arise for these intersections,
namely
%%%%%%
\be
\hbox{AdS$_3\times S^3$,
AdS$_3\times (S^2\times S^1)$,
(AdS$_2\times S^1)\times S^3$,
(AdS$_2\times S^1) \times (S^2\times S^1)$}\ .
\label{horizon}
\ee
%%%%%%%
(To be precise, the AdS$_3$ and the $S^3$ can in general be factored by
cyclic groups, in the manner discussed previously.)
If we oxidise these near-horizon solutions further, to $D=10$ or
$D=11$, then the additional dimensions provide additional factors of
$T^4$ or $T^5$ respectively.

For example, the four-charge black hole solution using the field
strengths of the last entry in the list (\ref{d4n4}) becomes an
intersection of $p$-branes in $D=6$, and its near-horizon region is
(AdS$_2\times S^1) \times (S^2\times S^1)$.  This is because the
Kaluza-Klein vector is not involved in the solution, and so the
six-dimensional metric is diagonal.  If the solution is built using
the set of field strengths in the first entry of the list
(\ref{d4n4}), two possibilities can arise.  If the Kaluza-Klein field
${\cal F}_\2^p$ carries a magnetic charge (implying that the other
three field strengths carry electric charges), then the solution
becomes (AdS$_2\times S^1)\times S^3$.  In other words, the
Kaluza-Klein vector describes a magnetic monopole which corresponds,
from the six-dimensional point of view, to a NUT charge that twists
the $S^2\times S^1$ product to give $S^3$.  On other hand, if the
Kaluza-Klein field carries an electric charge (and so the other three
field strengths carry magnetic charges), then the solution describes
AdS$_3\times (S^2\times S^1)$.  This is because in this case the
Kaluza-Klein vector has a configuration which, from the
six-dimensional point of view, corresponds to a wave which twists the
AdS$_2\times S^1$ product to give AdS$_3$.  The oxidation of the
solutions for the configurations of field strengths listed in the
second entry in (\ref{d4n4}) describe AdS$_3\times S^3$ in $D=6$.
This is because there are two Kaluza-Klein fields in this case, {\it
  viz}. ${\cal F}_\2^j$ and ${\cal F}_\2^k$.  One of them carries a
magnetic charge and hence twists the $S^2\times S^1$ product, while
the other carries an electric charge, and hence twists the
AdS$_2\times S^1$ product.

We showed in sections 3, 4 and 5 that the near-horizon structures
(\ref{horizon}) of the six-dimensional intersections that come from
the oxidations of the non-dilatonic black holes in $D=5$ and $D=4$ are
related to each other through Hopf T-duality.  Furthermore, even the
near-horizon limits of the 4-charge solutions which are, owing to sign
choices, non-supersymmetric, are related by Hopf T-duality to the
supersymmetric ones.  As we show in appendix D, the Hopf T-duality
not only relates the near-horizon limits listed in and below
(\ref{horizon}), but also relates the associated full solutions.  Thus
in particular, a Hopf reduction and T-duality has the effect of
mapping the solutions (\ref{d5n3}) and (\ref{d4n4}) among each other.
The harmonic function associated with the Kaluza-Klein vector coming
from the Hopf reduction lacks a constant term; however, it can be
introduced by performing an appropriate U-duality transformation
\cite{hyun,Boonstra1,cllpst,bb}.

\section*{Appendices}
\addcontentsline{toc}{part}{Appendices}
\appendix
\section{T-duality of the truncated six-dimensional theories}

In section 2, we obtained two different consistent truncations of
six-dimensional maximal supergravity.  One of them, given by
(\ref{2blag}), naturally arose as a consistent truncation of
six-dimensional maximal supergravity in the type IIB picture.  This
truncated theory has an $O(2,2)$ global symmetry.  The other theory,
given by (\ref{2alag}), naturally arose as a consistent truncation of
six-dimensional maximal supergravity in the type IIA picture.  This
theory has only an $\R \times \R$ global symmetry.  Interestingly, it
can be obtained as the dimensional reduction of the seven-dimensional
Lagrangian (\ref{7dlag}), which itself can easily be shown to be a
consistent truncation of seven-dimensional maximal supergravity.  In
this appendix, we shall show that the two different six-dimensional
Lagrangians (\ref{2blag}) and (\ref{2alag}) are T-dual to each other,
in the sense that upon dimensional reduction they give rise to the
same five-dimensional theory, up to field redefinitions.

     We begin by obtaining the dimensional reduction of the Lagrangian
(\ref{2blag}), coming from the consistent truncation in the type IIB
picture.  We find
%%%%%
\bea
e^{-1}\, {\cal L}_{5B} &=& R -\ft12 (\del\phi_1)^2
-\ft12(\del\phi_2)^2 -\ft12(\del\varphi)^2 -
\ft12 e^{2\phi_1}\, (\del\chi_1)^2
-\ft12 e^{2\phi_2}\, (\del \chi_2)^2 \nn\\
&&-\ft1{12} e^{-\phi_1 -\phi_2+\sqrt{\ft23}\varphi }\, (F_\3^\ns)^2
-\ft1{12} e^{\phi_1 -\phi_2+\sqrt{\ft23}\varphi }\, (F_\3^\rr)^2\nn\\
&&-\ft1{4} e^{-\phi_1 -\phi_2-\sqrt{\ft23}\varphi }\, (F_{\2 1}^\ns)^2
-\ft1{4} e^{\phi_1 -\phi_2-\sqrt{\ft23}\varphi }\, (F_{\2 1}^\rr)^2\nn\\
&&-\ft14 e^{\sqrt{\ft83}\varphi}\, (\cF_\2)^2
+ \chi_2\, dA_\2^\ns\wedge dA_{\1 1}^\rr-
\chi_2\, dA_\2^\rr\wedge dA_{\1 1}^\ns
\ .\label{2bd5lag}
\eea
%%%%%
Here, the field strengths are given by
%%%%%%
\bea
&&F_\3^\ns = dA_\2^\ns - dA^\ns_{\1 1}\wedge \cA_\1\ ,\qquad
F^\ns_{\2 1} = dA^\ns_{\1 1}\ ,\qquad
\cF_\2=d\cA_\1\ ,\label{d5bkk}\\
&&F_\3^\rr = dA_\2^\rr - dA^\rr_{\1 1} \cA_1 +
\chi_1(dA_\2^\ns - dA^\ns_{\1 1}\wedge \cA_\1)\ ,\quad
F^\rr_{\2 1} = dA^\rr_{\1 1} + \chi_1 dA^\ns_{\1 1}\ .\nn
\eea
%%%%%%%

     The five-dimensional Lagrangian coming from the dimensional
     reduction of the truncated theory (\ref{2alag}) is given by
%%%%%
\bea
e^{-1}\, {\cal L}_{5A} &=& R -\ft12 (\del\phi_1)^2
-\ft12(\del\phi_2)^2 -\ft12(\del\varphi)^2\nn\\
&&
-\ft12e^{-\ft12\phi_1 +\ft32\phi_2 -\sqrt{\ft32}\varphi} (\del\chi')^2
-\ft12 e^{\ft32\phi_1 -\ft12\phi_2 -\sqrt{\ft32}\varphi}
(\del A_{\0 1})^2\nn\\
&&
-\ft1{12} e^{-\phi_1-\phi_2+\sqrt{\ft23}\varphi}\, (F_\3)^2
-\ft1{12} e^{\fft12\phi_1-\fft32\phi_2-\ft1{\sqrt6}\varphi}\,
(F_{\3 1})^2\nn\\
&&
-\ft1{4} e^{-\phi_1-\phi_2-\sqrt{\ft23}\varphi}\, (F_{\2 1})^2
-\ft1{4} e^{\fft32\phi_1 -\fft12\phi_2 + \ft1{\sqrt6}\varphi}\, (F_\2)^2
\nn\\
&&
-\ft14 e^{\sqrt{\ft83} \varphi} (\cF_\2)^2 + \chi'\, F_{\3 1}\wedge
\cF_2 + \chi'\, F_\3\wedge F_\2\ ,\label{2ad5lag}
\eea
%%%%%
where we have dualised the 3-form potential $A_\3$ to an axion
$\chi'$.  The field strengths in (\ref{d5akk}) are given given by
%%%%%%%%%
\bea
&&
F_{\3 1}= dA_{\2 1} + dA_{\1 1}\wedge dA_\1 - A_{\0 1}\, dA_\2
\ ,\qquad F_{\2 1}=dA_{\1 1}\ ,\qquad
F_{\1 1} = dA_{\0 1}\ ,\nn\\
&&
F_{\3} = dA_\2 - dA_{\1 1}\wedge \cA_\1\ ,\qquad
F_{\2} = dA_\1 - dA_{\0 1} \wedge \cA_\1\ .\label{d5akk}
\eea
%%%%%%

     It can be verified that the two 5-dimensional Lagrangians
(\ref{2bd5lag}) and (\ref{2ad5lag}) are related to each other by the
field redefinition described in Table 1 and (\ref{lambda}).  To be
precise, we first make the following field redefinitions:
%%%%%%%%
\be
A_\1' = A_\1 - A_{\0 1}\,\cA_\1\ ,\qquad
A_\2'= A_\2 - A_{\1 1} \wedge \cA_\1\ ,\qquad
A_{\2 1}'= A_{\2 1} + A_{\1 1}\wedge A_\1'\ .
\ee
%%%%%%%
After doing this we find that the Lagrangian (\ref{2ad5lag}) can be
mapped to the Lagrangian (\ref{2bd5lag}) by the following
transformations
%%%%%%%%%
\bea
&&
A_\2'\longrightarrow A_\2^\ns\ ,
\quad A_{\2 1}' \longrightarrow - A_\2^\rr\ ,\quad
\cA_1\longrightarrow A_{\1 1}^\ns\ ,
\quad A_{\1}' \longrightarrow A_{\1 1}^\rr\ ,\nn\\
&&
A_{\1 1}\longrightarrow \cA_\1\ ,
\quad A_{\0 1} \longrightarrow \chi_1\ ,\quad
\chi' \longrightarrow \chi_2\ ,
\eea
%%%%%%%%%
together with the transformation of the dilatons given by
(\ref{lambda}).

\section{$O(2,2)$ symmetry of the truncated six-dimensional theory}

    Here, we give the explicit $O(2,2) \sim SL(2,\R)_1\times
SL(2,\R)_2$ global symmetry transformations for the truncated
six-dimensional theory (\ref{2blag}).  The factor $SL(2,\R)_1$ is
an S-duality symmetry that maps between the NS-NS and R-R 2-form
potentials, and is realised at the level of the Lagrangian.  The
factor $SL(2,\R)_2$ is an electric/magnetic duality symmetry between
the NS-NS and R-R 3-form field strengths, which is realised only at
the level of the equations of motion.

   To present the global transformation rules, it is useful first to
define the two complex scalar fields
%%%%%
\be
\tau_1 \equiv \chi_1 + \im\, e^{-\phi_1}\ ,\qquad
\tau_2 \equiv \chi_2 + \im\, e^{-\phi_2}\ .
\ee
%%%%%
The two $SL(2,\R)$ transformations act non-linearly on the scalar
manifold as follows:
%%%%%
\bea
SL(2,\R)_1: && \tau_1 \longrightarrow \Lambda_1\cdot \tau_1\equiv
             \fft{a_1\, \tau_1 + b_1}{c_1\, \tau_1 + d_1}\ ,\qquad
                \tau_2\longrightarrow \tau_2\ ,\nn\\
SL(2,\R)_2: && \tau_2 \longrightarrow \Lambda_2\cdot\tau_2\equiv
             \fft{a_2\, \tau_2 + b_2}{c_2\, \tau_2 + d_2}\ ,\qquad
                \tau_1\longrightarrow \tau_1\ ,\label{sl2rscalars}
\eea
%%%%%
where $a_1\, d_1-b_1\, c_1 =1 = a_2\, d_2 - b_2\, c_2$, and we may
define the $SL(2,\R)$ matrices $\Lambda_1$ and $\Lambda_2$ in the
standard way:
%%%%%
\be
\Lambda_1 = \pmatrix{ a_1 & b_1\cr
                      c_1 & d_1}\ ,\qquad
\Lambda_2 = \pmatrix{ a_2 & b_2\cr
                      c_2 & d_2}\ .
\ee
%%%%%

    $SL(2,\R)_1$ is a symmetry of the Lagrangian, and it acts linearly
on the 2-form potentials:
%%%%%
\be
A_\2 \equiv \pmatrix{A_\2^\ns \cr A_\2^\rr} \longrightarrow
         (\Lambda_1^T)^{-1}\, A_\2 \ .
\ee
%%%%%
$SL(2,\R)_2$ is a symmetry only at the level of the equations of
motion, and it acts locally only on the field strengths.  We shall
also, for convenience, present the action of $SL(2,\R)_1$ on the field
strengths.  We begin by defining two field-strength doublets,
$H_\3^1$ and $H_\3^2$:
%%%%%
\be
H_\3^1 = \pmatrix{ e^{-\phi_1}\, F_\3^\ns + \chi_1\, e^{\phi_1}\,
  F_\3^\rr \cr e^{\phi_1}\, F_\3^\rr} \ ,\qquad
H_\3^2 =\pmatrix{ e^{-\phi_2}\, {*F_\3^\ns} + \chi_2\,
  F_\3^\rr \cr  F_\3^\rr}\ .\label{hdef}
\ee
%%%%%
Their $SL(2,\R)$ transformations are:
%%%%%
\bea
SL(2,\R)_1: && H_\3^1\, \longrightarrow \Lambda_1\, H_\3^1\ ,\nn\\
SL(2,\R)_2: && H_\3^2\, \longrightarrow \Lambda_2\, H_\3^2\ .\label{sl2rh}
\eea
%%%%%
Note that the two fields $H_\3^1$ and $H_\3^2$ are not independent,
and so given the transformation on one, the transformation on the
other is in principle determined.  Each was introduced for the
specific purpose of encoding one or other of the two $SL(2,\R)$
transformations in a simple way, as indicated in (\ref{sl2rh}), and we
do not need to give the associated transformations on the other $H_\3$
field.

    The theory can describe strings that carry four independent
charges, namely the electric and the magnetic charges for both the
NS-NS and the R-R 3-forms.  It is easily seen from the equations of
motion following from (\ref{2blag}), and from the Bianchi identities
for $F_\3^\ns$ and $F_\3^\rr$,  that they are given by
%%%%%
\bea
Q_e^\ns &=& \ft1{16\pi^2}
        \int\Big\{e^{-\phi_1-\phi_2}\, {* F_\3^\ns} + \chi_2\,
  F_\3^\rr + \chi_1\, \Big(e^{\phi_1-\phi_2}\, {*F_\3^\rr} -\chi_2\,
  F_\3^\ns\Big) \Big\}\ ,\nn\\
Q_m^\ns &=& \ft1{16\pi^2} \int F_\3^\ns\ ,\nn\\
Q_e^\rr &=& \ft1{16\pi^2} \int
          \Big\{ e^{\phi_1-\phi_2}\, {*F_\3^\rr} -\chi_2\,
F_\3^\ns \Big\}\ ,\nn\\
Q_m^\rr &=&\ft1{16\pi^2} \int \Big\{ F_\3^\rr -\chi_1\, F_\3^\ns\Big\}
\ .\label{o22charges}
\eea
%%%%%
>From the $SL(2,\R)_1$ and $SL(2,\R)_2$ transformations rules
(\ref{sl2rscalars}) and (\ref{sl2rh}), we find that these charges
transform as:
%%%%%
\bea
SL(2,\R)_1: && \vec Q_{ee}\equiv
                       \pmatrix{Q_e^\ns \cr Q_e^\rr} \longrightarrow
               \Lambda_1\, \vec Q_{ee}\ ,\quad
\vec Q_{mm}\equiv \pmatrix{Q_m^\ns \cr Q_m^\rr} \longrightarrow
               (\Lambda_1^T)^{-1}\, \vec Q_{mm}\ ,\nn\\
SL(2,\R)_2: && \vec Q_{em}\equiv
                   \pmatrix{Q_e^\ns \cr Q_m^\rr} \longrightarrow
               \Lambda_2\, \vec Q_{em}\ ,\quad
\vec Q_{me}\equiv \pmatrix{Q_m^\ns \cr Q_e^\rr} \longrightarrow
               (\Lambda_2^T)^{-1}\,  \vec Q_{me}\ .\label{sl2rcharges}
\eea
%%%%%
Here we are introducing the notation that $\vec Q_{xy}$ is a
two-component charge vector, whose upper component is the electric
($x=e$) or magnetic ($x=m$) NS-NS charge, and whose lower component is
the electric ($y=e$) or magnetic ($y=m$) R-R charge.

It is worthwhile pausing at this point, to understand why the charge
vectors $\vec Q_{mm}$ and $\vec Q_{me}$ transform contragrediently in
comparison to the transformations of $\vec Q_{ee}$ and $\vec Q_{em}$.
If we introduce an index notation for the two fields, so that $F_\3^i$
denotes the NS-NS field when $i=1$, and the R-R field when $i=2$, then
the dual fields $\tF_{\3 1}= e^{-\phi_1-\phi_2}\, {*F_\3^1}$ and
$\tF_{\3 2}= e^{\phi_1-\phi_2}\, {*F_\3^2}$ actually correspond to the
first and second components of a doublet $\tF_{\3 i}$ with a {\it
  downstairs} index $i$.  (This can be seen from the fact that the
relevant $SL(2,\R)_1$-invariant kinetic terms in the Lagrangian have
the form $-\ft12 \tF_{\3 i}\wedge F_\3^i$ \cite{cjlp2}.)  The electric
charges therefore can be labelled as $Q_{e\, i}=\{Q_e^\ns, Q_e^\rr\}$,
while the magnetic charges can be labelled as
$Q_m^i=\{Q_m^\ns,Q_m^\rr\}$.  Since the electric charges transform
directly with $\Lambda$, we can see that the suppressed matrix indices
on $\Lambda_1$, if made explicit, are located as follows:
$(\Lambda_1)_i{}^j$.  It is now clear why it is the contragedient
representation that acts on the magnetic charges $Q_m^i$.  An
analogous comment applies to the electric and magnetic indices
associated with the $SL(2,\R)_2$ transformations.

   It is convenient to give a $4\times4$ matrix representation for the
$SL(2,\R)_1\times SL(2,\R)_2$ transformations, by defining the tensor
product of $2\times 2$ matrices, and 2-component vectors,  as follows:
%%%%%
\bea
M\otimes N &\equiv& \pmatrix{ m_{11}\, N & m_{12}\, N\cr
                           m_{21}\, N & m_{22}\, N}\ ,\nn\\
&&\nn\\
\pmatrix{ x\cr y} \otimes \pmatrix{ u\cr v} &\equiv&
\pmatrix{ x\, u\cr
                     x\, v\cr y\, u \cr y\, v}\ ,\label{tensorprod}
\eea
%%%%%
where $m_{ij}$ denotes the components of $M$. In view of the remarks
in the previous paragraph, it follows that we can represent the action
of the two $SL(2,\R)$'s on the charges as $\vec Q \longrightarrow
\Lambda\, \vec Q$, where $\Lambda = \Lambda_2\otimes \Lambda_1$, and
%%%%%
\be
\vec Q = \pmatrix{Q_e^\ns\cr Q_e^\rr \cr Q_m^\rr \cr -Q_m^\ns}\ .
\label{4chargevec}
\ee
%%%%%

    For later convenience, we may also now introduce an alternative
parameterisation of the scalar coset manifold.  It is sufficient for
this purpose to consider a generic $SL(2,\R)/O(2)$ coset, and the
formalism can then be applied to both $SL(2,\R)_1$ and $SL(2,\R)_2$.
We define the upper-triangular matrix
%%%%%
\be
\v =\pmatrix{ e^{-\fft12\phi} & \chi\, e^{\fft12\phi}\cr
                   0 & e^{\fft12\phi}}\ .
\ee
%%%%%
This gives a Borel parameterisation of the coset, whose Lagrangian can
now be written as $\ft14 \tr (\del_\mu({\cal M}^{-1})\, \del^\mu{\cal
M})$, where
%%%%%
\be
{\cal M} = \v\, \v^T=
               \pmatrix{e^{-\phi} + \chi^2\, e^\phi & \chi\, e^\phi\cr
                    \chi\, e^\phi & e^\phi}\ .
\ee
%%%%%
The $SL(2,\R)$ transformation (\ref{sl2rscalars}) of the scalar fields
can then be expressed as $\v\longrightarrow \Lambda\, \v\, {\cal O}$,
where ${\cal O}$ is a field-dependent compensating $O(2)$
transformation that restores $\v$ to the Borel gauge.  On the matrix
${\cal M}$, the $SL(2,\R)$ transformation is simply ${\cal M}
\longrightarrow \Lambda\, {\cal M}\, \Lambda^T$.  Note that $H_\3^1$
in (\ref{hdef}) can now be written as $H_\3^1={\cal M}_1\, dA_\2$.

With these preliminaries, we are now in a position to construct, in
the next section, an $O(2,2)$ multiplet of dyonic strings

\section{An $O(2,2;\Z)$ multiplet of dyonic strings}

   We may construct general dyonic string solutions with arbitrary
electric and magnetic NS-NS and R-R charges $(Q_e^\ns, Q_m^\ns,
Q_e^\rr, Q_m^\rr)$ at an arbitrary modulus point specified by
$\tau_1^0$ and $\tau_2^0$, by starting with the simple case of an NS-NS
dyonic string in the $\tau_1^0=\tau_2^0=\im$ vacuum, with electric and
magnetic NS-NS charges $q$ and $p$.  We then act with $O(2)_1$ and
$O(2)_2$ transformations
%%%%%
\be
\Lambda_1(\theta_1) =\pmatrix{\cos\theta_1 & \sin\theta_1\cr
                    -\sin\theta_1 & \cos\theta_1}\ ,\qquad
\Lambda_2(\theta_2) =\pmatrix{\cos\theta_2 & \sin\theta_2\cr
                    -\sin\theta_2 & \cos\theta_2}\ ,\label{o2trans}
\ee
%%%%%
which lie in the stability subgroups of the $SL(2,\R)$'s that rotate
the charges while leaving the $\tau^0=\im$ modulus points fixed.
Next, we act with the Borel transformations
%%%%%
\be
\Lambda_1(\tau_1^0) = \pmatrix{ e^{-\fft12\phi^0_1} & \chi^0_1\,
                                             e^{\fft12\phi^0_1}\cr
                   0 & e^{\fft12\phi^0_1}}\ ,\qquad
\Lambda_2(\tau_2^0) = \pmatrix{ e^{-\fft12\phi^0_2} & \chi^0_2\,
                                             e^{\fft12\phi^0_2}\cr
                   0 & e^{\fft12\phi^0_2}}\ ,\label{boreltrans}
\ee
%%%%%
which map the original modulus points $\tau_1^0=\tau_2^0=\im$ to the
arbitrary points $\tau_1^0=\chi_1^0 + \im\, e^{-\phi_1^0}$ and
$\tau_2^0=\chi_2^0 + \im\, e^{-\phi_2^0}$. The combined effect of the
stability-subgroup and Borel-subgroup transformations is to map the
original charges $\vec Q= (Q_e^\ns, Q_e^\rr, Q_m^\rr, -Q_m^\ns) =
(q,0,0,-p)$ to arbitrary charges, which are related to the parameters
$\theta_1$, $\theta_2$, $q$ and $p$ (at the given modulus point
$(\phi_1^0, \chi_1^0, \phi_2^0, \chi_2^0)$) in a manner that we shall
determine below.  Thus we can obtain strings with their four charges
lying at arbitrary points on the charge lattice (that satisfy the
Dirac quantisation condition) by appropriately choosing the four
parameters.  (The spirit of this construction is similar to that used
in \cite{duffrahm} in the discussion of the charge lattice for the
$O(6,22;\Z)$ multiplet of black holes in the $D=4$ heterotic string,
and in \cite{schwarz} for the construction of the $SL(2,\Z)$ multiplet
of type IIB strings.  A general procedure for generating the U-duality
multiplets for all $p$-brane solitons was given in \cite{trombone}.  A
group theoretic approach was also introduced in \cite{trombone}, using
the homogeneous scaling symmetries of the equations of motion that
arise in theories such as the maximal supergravities, in order to give
a construction of genuine spectrum-generating groups for BPS states.
An explicit construction of U-duality multiplets for BPS states in
eight dimensions was given in \cite{lr}.)

\subsection{Unboosted isotropic dyonic strings}

    Moving now to the details, let us consider an isotropic unboosted
dyonic string, supported by the NS-NS 3-form field \cite{Rahmfeld};
%%%%%
\bea
ds^2 &=& (H_e\, H_m)^{-1/2}\, dx^\mu\, dx_\mu + (H_e\, H_m)^{1/2}\,
         d\vec y\cdot d\vec y\ ,\nn\\
F_\3^\ns &=& 8\, p\, \Omega_\3 + 8\, q\, H_m \, H_e^{-1} \, {*Omega_\3}\ ,
 \qquad  F_\3^\rr=0\ ,\nn\\
\tau_1&=& \chi_1 + \im\, e^{-\phi_1} =
\tau_2 =\chi_2 + \im\, e^{-\phi_2}= \im\, (H_e/H_m)^{1/2}\ ,\label{nsdyonic}
\eea
%%%%%
where $H_e=1+4q/r^2$ and $H_m=1+4p/r^2$, $q$ and $p$ are the electric
and magnetic charges (following the normalisations given in
(\ref{o22charges})), and $\Omega_\3$ is the volume form on the unit
$S^3$.  After some algebra, we find, following the steps outlined above, that
the solution after performing the stability-subgroup and
Borel-subgroup transformations becomes
%%%%%
\bea
ds^2 &=& (H_e\, H_m)^{-1/2}\, dx^\mu\, dx_\mu + (H_e\, H_m)^{1/2}\,
         d\vec y\cdot d\vec y\ ,\nn\\
\tau_1 &=& \chi_1^0 +\im\, e^{-\phi_1^0}\,
\fft{\sqrt{H_e} - \im\, \tan\theta_1\,
  \sqrt{H_m}}{\sqrt{H_m} -\im \, \tan\theta_1\, \sqrt{H_e}}\ ,\nn\\
\tau_2 &=& \chi_2^0 +\im\, e^{-\phi_2^0}\,
 \fft{\sqrt{H_e} - \im\, \tan\theta_2\,
  \sqrt{H_m}}{\sqrt{H_m}-\im \tan\theta_2\, \sqrt{H_e}}\ ,
     \label{o22dyonic}\\
F_\3^\ns &=&e^{\fft12(\phi_1^0+\phi_2^0)}\, \Big( \cos\theta_1\,
\cos\theta_2 \, \Theta - \sin\theta_1\, \sin\theta_2\,
\fft{H_e}{H_m}\, {*\Theta} \Big)\ , \nn\\
F_\3^\rr &=& \fft{-e^{\ft12(\phi_2^0-\phi_1^0)}\, H_e}{
       \sin^2\theta_1\, H_e + \cos^2\theta_1\, H_m}\, \Big(
\sin\theta_1\, \cos\theta_2\, \Theta + \cos\theta_1\, \sin\theta_2\,
{*\Theta}\Big)\ ,\nn
\eea
%%%%%
where $\Theta\equiv 8\, p\, \Omega_\3 + 8\,q\, {*\Omega_\3}$.

    Under the stability-subgroup and Borel-subgroup transformations
(\ref{o2trans}) and (\ref{boreltrans}), the initial charge 4-vector
$\vec Q_0 = \{q,0,0,-p\}$ is mapped to the final charge vector $\vec
Q_f$, given by
%%%%%
\be
\vec Q_f = \Big(\Lambda_2(\tau_2^0)\otimes
\Lambda_1(\tau_1^0)\Big)\, \Big(\Lambda_2(\theta_2)\otimes
\Lambda_1(\theta_1)\Big)\, \vec Q_0\ ,
\ee
%%%%%
which implies
%%%%%
\be
\Big(\Lambda_2(\theta_2)\otimes
\Lambda_1(\theta_1)\Big)\, \vec Q_0=
\Big(\Lambda_2(\tau_2^0)^{-1}\otimes
\Lambda_1(\tau_1^0)^{-1}\Big)\,\vec Q_f \equiv \Lambda(\tau_0)^{-1}\,
\vec Q_f\ .\label{4rel}
\ee
%%%%%
This equation provides the relation between the four parameters
$(\theta_1,\theta_2,q,p)$ and the four final charges $\vec Q_f$, for
any given values of the scalar moduli $\tau_0=(\tau_1^0,\tau_2^0)$. To
obtain the explicit solution, we first note that the stability-subgroup
$O(2)$ rotations (\ref{o2trans}) can be written as $
\Lambda(\theta)= e^{\im\theta \sigma}$,
where $\sigma=\pmatrix{0&-\im\cr\im &0}$.  Equation (\ref{4rel}) can
now be written as
%%%%%
\be
e^{\im\theta_2\sigma} \otimes e^{\im\theta_1\sigma} \, \vec Q_0
=\Lambda(\tau_0)^{-1}\, \vec Q_f\ .
\ee
%%%%%
This implies that $e^{\im\theta_2\sigma} \otimes e^{\im\theta_1
\sigma} \,U_\pm = V_\pm$, where
%%%%%
\be
U_\pm = (\oneone+\sigma)\otimes (\oneone\pm\sigma)\, \vec Q_0\ , \qquad
V_\pm =(\oneone+\sigma)\otimes (\oneone\pm\sigma)\, \vec Q_f\ .
\ee
%%%%%
After elementary algebra, it now follows that
%%%%%
\bea
(q+p)\, e^{\im(\theta_2+\theta_1)} &=& e^{\fft12(\phi_1^0+\phi_2^0)}\,
(Q_e^\ns - \tau_1^0\, Q_e^\rr - \tau_2^0\, Q_m^\rr - \tau_1^0\,
\tau_2^0\, Q_e^\ns)\equiv \Delta_+\ ,\nn\\
(q-p)\, e^{\im(\theta_2-\theta_1)} &=& e^{\fft12(\phi_1^0+\phi_2^0)}\,
(Q_e^\ns - \bar\tau_1^0\, Q_e^\rr - \tau_2^0\, Q_m^\rr - \bar\tau_1^0\,
\tau_2^0\, Q_e^\ns)\equiv \Delta_-\ .\label{4relans}
\eea
%%%%%
>From here, the solutions for $(\theta_1, \theta_2,q,p)$ immediately
follow.  Note that the factor $e^{\fft12(\phi_1^0+\phi_2^0)}$ is
precisely the six-dimensional effective string coupling constant.

It is now easy to determine the formula for the mass $m$ per unit
length for the general 4-charge dyonic string.  To do this, we note
that in the original NS-NS dyonic string solution (\ref{nsdyonic}),
the mass is simply given by $m = q+p$, which we can write in the
$O(2,2)$-invariant form
%%%%%
\be
m^2 = q^2 + p^2 + 2 q\, p = \vec Q_0^T \, \vec Q_0 - \vec Q_0^T\,
\hat \Omega\, \vec Q_0\ ,\label{mass1}
\ee
%%%%%
where $\hat\Omega \equiv \Omega\otimes \Omega$, and
$\Omega=\pmatrix{0&1\cr -1 &0}$.  In terms of the
final charges $\vec Q_f$ of the generic 4-charge dyonic string, the
mass is therefore given by
%%%%%
\bea
m^2 &=& \vec Q_f^T \, (\Lambda(\tau^0)^T)^{-1}\,
\Lambda(\tau^0)^{-1}\, \vec Q_0 - \vec Q_f^T  \, (\Lambda(\tau^0)^T)^{-1}\,
\hat\Omega\, \Lambda(\tau^0)^{-1}\, \vec Q_0\nn\\
&=& \vec Q_f^T\, \Big((\Lambda(\tau^0)\, \Lambda(\tau^0)^T)^{-1} -
\hat\Omega\Big)\, \vec Q_f\ .\nn\\
&=& |\Delta_+|^2\ .\label{o22mass}
\eea
%%%%%
(The expression for $m^2$ in the last line follows directly from
(\ref{mass1}) and (\ref{4relans}).)

     Note that the second line of the mass formula (\ref{o22mass}) is
composed of two independent $O(2,2)$-invariant quantities.  The second
term is precisely the quantity that appears in the Dirac quantisation
condition \cite{blps}, namely
%%%%%
\be
\vec Q_f^T\, \hat\Omega\, \vec Q_f = \hbox{integer}\ .
\ee
%%%%%
This condition implies that the four charges $\vec Q_f =(Q_e^\ns,
Q_e^\rr, Q_m^\rr, -Q_m^\ns)$ must lie on a discrete lattice, which,
for simplicity, may be taken to be the square integer lattice.  In our
construction of the multiplet of integer-charge solutions, we of
course allowed the initial charge parameters $q$ and $p$ and the
rotation angles $\theta_1$ and $\theta_2$ to be unrestricted by any
quantisation condition.  For a given modulus point
$\tau^0=(\tau_1^0,\tau_2^0)$, after restricting the final charges to
lie on the Dirac charge lattice, the initial parameters will
themselves fill out only a discrete lattice of values.

The points in a given charge lattice, for example the integer square
lattice, can be filled out by a discrete $O(2,2;\Z)$
spectrum-generating group that leaves the scalar modulus point
$\tau^0$ invariant.  It is therefore not itself the discretised form
of the of the $O(2,2)$ global symmetry transformations that we
originally discussed, since this transforms the scalar moduli at the
same time as it moves the charges on their lattice.  The difference is
highlighted by the fact that the spectrum-generating group must, of
course, in particular generate strings of different masses, whilst the
original global symmetry group leaves the metric, and hence the mass,
invariant.  This issue was extensively discussed in \cite{trombone},
where it was shown that to get the true spectrum-generating group it
is necessary to make use also of an homogeneous scaling ``trombone''
symmetry of the theory.  In \cite{trombone}, the example of the
$SL(2,\Z)$ multiplet of type IIB strings was discussed in detail.
However, in the $O(2,2)$ case we are considering here there is an
added subtlety.  This can be seen most easily at the classical level.
In the case of the type IIB NS-NS and R-R strings, the charge space is
two-dimensional, and can be completely spanned, for any given modulus
point, by the action of the $O(2)$ denominator group together with the
trombone rescaling symmetry. In our present case, however, the charge
space is four-dimensional, while the modulus-preserving denominator
group is $O(2)\times O(2)$.  Together with the trombone symmetry this
gives only three parameters, and hence this is insufficient to span
the charge vector space.  In our construction, we were nevertheless
able to construct the complete charge lattice of dyonic strings. This
is because we started with the solution (\ref{nsdyonic}) that had the
two free parameters $q$ and $p$.  The trombone symmetry is responsible
for the existence of the 1-parameter sub-family of solutions where $q$
and $p$ are uniformly rescaled by the same factor, which has the
effect of also rescaling the mass.  It is less clear what is the
symmetry that is responsible for allowing solutions with different
relative ratios between the electric and magnetic charge parameters
$q$ and $p$.

     It was argued in \cite{trombone} that one resolution to this
puzzle might be that since the dyonic string can be viewed as a bound
state with {\it zero} binding energy, it is less fundamental than the
individual electric and magnetic building blocks.   Thus it would be
unnecessary to find a symmetry to relate the bound-state solutions with
different ratios for the electric and magnetic charge.  (Another
example where there is no homogeneous scaling symmetry  to account for
the existence of arbitrary-charge solutions is in the heterotic
string.  This is because Yang-Mills fields $F\sim dA + A\wedge A$
do not scale uniformly under $A\rightarrow \lambda \, A$.  (One
requires a scaling symmetry that rescales the metric, so that the mass
can be changed, while leaving the scalars invariant, so that the
scalar moduli are unchanged.))

    In this section, we constructed an $O(2,2;\Z)$ multiplet of dyonic
strings in $D=6$.  This subgroup of the full $O(5,5;\Z)$ or
$O(5,21;\Z)$ duality groups captures the essence of the complete
groups, in that it describes both the NS-NS/R-R duality and the
electric/magnetic duality.  The extension to the full duality group is
straightfoward in principle, using, for example, the techniques
described in \cite{trombone}, but its detailed implementation would be
tedious and rather unrewarding.  The most interesting aspect of the
U-duality multiplets is to see how the charges transform under the
full U-duality group.  The classification of the U-duality orbits of
the charges in all dimensions can be found in \cite{lpsorbit,fg}.

\subsection{Boosted and twisted dyonic strings}

In the discussion above, we gave the construction of the $O(2,2;\Z)$
multiplet of unboosted, isotropic dyonic strings.  It is
straightforward to boost in the worldvolume, or to twist in the
transverse space, and thereby obtain $O(2,2;\Z)$ multiplets of boosted
or twisted dyonic strings.  We do this by following a strategy analogous to
the one we used previously, namely by starting with a boosted and
twisted dyonic string supported purely by NS-NS 3-form charges, and
taking the scalar moduli to vanish. The form of the solutions for the
fields $(\phi_1,\chi_1,\phi_2,\chi_2, F_\3^\ns, F_\3^\rr)$ is
identical to that for the unboosted, untwisted case given in
(\ref{nsdyonic}), except that now the harmonic functions $H_e$ and
$H_m$ are modified to $H_e=1 + q/r$, $H_m=1+p/r$. (In other words,
they are now harmonic only in the 3-dimensional overall transverse
space, rather than the 4-dimensional transverse space for the dyonic
strings.\footnote{Recall that we are defining charges arising from
integrals over $S^2$ and over $S^3$ by $Q=\ft1{4\pi}\int F_2$ and
$Q=\ft1{16\pi^2}\int F_3$ respectively, which ensures that the charge
is preserved under dimensional reduction.  It follows that the harmonic
functions in the two cases will be of the form $H=1+Q/r$ and
$H=1+4Q/r^2$ respectively.})

 The metric is given by
%%%%%%%%%%
\bea
ds^2 &=& -(H_eH_m)^{-\ft12}\, K_w^{-1}\, dt^2 +
(H_eH_m)^{-\ft12}\, K_w\, (dz_1+(K_w^{-1}-1)dt)^2\nn\\
&&+(H_eH_m)^{\ft12}\, K_N^{-1}\, (dz_2 + Q_N\, \cos\theta\,d\phi)^2
\nn\\
&&+ (H_eH_m)^{\ft12}\, K_N\, (dr^2 + r^2d\theta^2 + r^2\sin^2\theta\,
d\phi^2) \ ,\label{btdyon}
\eea
%%%%%%%%%%%
where the extra harmonic functions for the wave and the boost charges
are $K_w=1+Q_w/r$ and $K_N=1+ Q_N/r$.  We may now repeat the steps
that we followed previously, to generate the entire $O(2,2;\Z)$
multiplet of boosted and twisted dyonic strings.  The final
expressions for the dilatons, axions and 3-form fields will be
identical to those given in (\ref{o22dyonic}), again with the
understanding that the $H_e$ and $H_m$ harmonic functions are modified
to those given above.  The reason for this is that the only other
change in the starting point is in the metric, which is a singlet
under $O(2,2)$, and so the entire calculation of the $O(2,2;\Z)$
multiplet proceeds identically to the one we described previously.
Note that the charges $p$ and $q$ are still given by (\ref{4relans}),
as in the case of unboosted isotropic dyonic strings.

     The solution (\ref{btdyon}) can be dimensionally reduced on the
fibre coordinates $z_1$ and $z_2$, giving rise to a non-dilatonic
black hole in $D=4$.   The near-horizon limit in $D=4$ is
AdS$_2\times S^2$, and so the area of the event horizon is
non-vanishing, implying that there is a non-vanishing entropy, even
though the solution is extremal.  The entropy is of the form
%%%%%
\be
S\sim \sqrt{p\, q\, Q_w\, Q_N}\ .\label{ent}
\ee
%%%%%
>From the six-dimensional point of view, the near-horizon limit is also
locally non-singular, and is given, after a rescaling of the time
coordinate, by
%%%%%
\bea
ds^2 &=& \sqrt{p\, q}\, Q_N\, \Big\{-e^{2\rho}\, dt^2 + d\rho^2
 +(\sqrt{\fft{Q_w}{p\, q\, Q_N}}\, dz_1 + e^\rho\, dt)^2 \nn\\
&&+ d\theta^2 + \sin^2\theta\, d\phi^2
+ (Q_N^{-1}\, dz_2 + \cos\theta\,
       d\phi)^2 \Big\}\ .\label{nh6}
\eea
%%%%%
It is straightforward to verify that the 4-volume of the spatial
metric at $\rho=-\infty$ is of the form (\ref{ent}).  Note that $Q_w$
measures the momentum of the wave propagating on the world-sheet of
the dyonic string, and it, together with the other charges, has the
effect of rescaling the fibre coordinate $z_1$.  $Q_N$, on
the other hand, is the NUT charge, and it has the effect of changing
the local structure of the $S^3$ factor to the lens space
$S^3/Z_{Q_N}$.  Thus we see that the phenomenon of $S^3$ being
factored to become a cyclic lens space plays a r\^ole in the
understanding of the entropy of 4-dimensional black holes.

If we consider the case where the dyonic string is only twisted, but
not boosted, the solution is given by (\ref{btdyon}) with $Q_w=0$. In
this case a dimensional reduction on the fibre coordinate $z_2$ gives
rise to a non-dilatonic string in $D=5$, which is dual to a
non-dilatonic 3-charge black hole. If, on the other hand, we instead
consider a case where there is only boosting, but no twisting, of the
dyonic string, the solution is given by
%%%%%
\bea
ds^2 &=& -(H_eH_m)^{-\ft12}\, K_w^{-1}\, dt^2 +
(H_eH_m)^{-\ft12}\, K_w\, (dz_1+(K_w^{-1}-1)dt)^2\nn\\
&&+(H_eH_m)^{\ft12}\,(dr^2 + r^2\,  d\Omega_3^2)\ ,\label{emw6}
\eea
%%%%%
where now the functions $H_e$, $H_m$ and $K_w$ are harmonic in a
4-dimensional transverse space, and so $H_e=1+4q/r^2$, $H_m=1+4p/r^2$ and
$K_w = 1 + 4Q_w/r^2$.  Its dimensional reduction on the fibre
coordinate $z_1$ gives a non-dilatonic 3-charge black hole in $D=5$.
The near-horizon limit of the six-dimensional metric (\ref{emw6}),
after rescaling the time coordinate, is
%%%%%
\bea
ds^2 &=& \sqrt{p\, q}\, \Big\{-e^{2\rho}\, dt^2 + d\rho^2
 +(\sqrt{\fft{Q_w}{p\, q}}\, dz_1 + e^\rho\, dt)^2 \nn\\
&&\qquad\qquad + d\theta^2 + \sin^2\theta\, d\phi^2
+ (dz_2 + \cos\theta\,
       d\phi)^2 \Big\}\ .\label{nh62}
\eea
%%%%%
The entropy is of the form $S\sim \sqrt{p\, q\, Q_w}$.

\section{Hopf reductions and isentropic mappings}

We have seen already that the Hopf T-duality transformations on the
fibre coordinates of AdS$_3$ or $S^3$, which can have the effect of
(un)twisting and squashing the metrics, preserve the entropy of the
solutions.  For example, we observe that the near-horizon limit
(\ref{nh62}) is the same as (\ref{nh6}) with $Q_N=1$, in which case
the lens space $S^3/Z_{Q_N}$ just becomes the 3-sphere.  The entropy
is therefore given by setting $Q_N=1$ in (\ref{ent}).  This shows,
from a six-dimensional point of view, that the near-horizon limit of a
$D=5$ isotropic 3-charge black hole is the same as the
near-horizon limit of a 4-charge black hole in $D=4$, where the
magnetic Kaluza-Klein charge $Q_N$ is set to 1.  In other words, in
the Hopf reduction on the $U(1)$ fibres of the 3-spheres that foliate
the transverse space of the 3-charge black hole in $D=5$, a fourth
charge ($Q_N=1)$ emerges in $D=4$, corresponding to the magnetic
charge of the Kaluza-Klein 2-form that governs the curvature of the
fibre bundle.  This should be contrasted with the more common kind of
Kaluza-Klein reduction of the 3-charge black hole in $D=5$, where the
reduction is on one of the coordinates of the transverse space.  In
this situation, it is is necessary first to make a line (or periodic
array) of $D=5$ black holes along the intended reduction axis.  The
consequence of this is that the number of charges is conserved under
the reduction process, and thus one arrives at a 3-charge black hole
in $D=4$ which is singular on the horizon, and which has zero entropy.
Hopf reduction, however, preserves the area of the horizon, and
provides a natural ``isentropic mapping'' between the $D=5$, 3-charge
and $D=4$, 4-charge black holes.

   To make this property of the Hopf reductions more explicit, let us
consider the entire solution for a 3-charge black hole in $D=3$,
rather than just looking at the near-horizon limit.  The 3-charge
solution, which can be obtained, for example, from the dimensional
reduction of (\ref{emw6}) on the $z_1$ coordinate, is
%%%%%
\be
ds_5^2 = -(H_e H_m K_w)^{-\ft23}\, dt^2 + (H_e H_m K_w)^{\ft13}\,
(dr^2 + r^2\, d\Omega_3^2)\ ,
\ee
%%%%%
where
%%%%%
\be
H_e= 1 + \fft{4q}{r^2}\ , \qquad H_m = 1+\fft{4p}{r^2}\ ,\qquad
K_w = 1 + \fft{4Q_w}{r^2}\ .\label{d5harm}
\ee
%%%%%
Writing the unit 3-sphere metric $d\Omega_3^2$ in the fibre bundle
form (\ref{fib32}), we now perform the reduction on the fibre
coordinate $z$, using the usual Kaluza-Klein ansatz $ds_5^2 =
e^{-\varphi/\sqrt3}\, ds_4^2 + e^{2\varphi/\sqrt3}\, (dz+B)^2$, giving
%%%%%
\be
ds_4^2 = -\ft12 r\, (H_e H_m K_w)^{-\ft12} \, dt^2 + \ft12 r\, (H_e
H_m K_w)^{\ft12} \, (dr^2 + \ft14 r^2\, d\Omega_2^2)\ .\label{d41}
\ee
%%%%%
The new Kaluza-Klein potential $B$ is such that ${\cal F}_\2=
dB=\Omega_\2$, and hence the Kaluza-Klein magnetic charge is
$Q_N=\ft1{4\pi} \int {\cal F}_\2 = 1$.

   This is not quite like a normal black hole solution in $D=4$ for two
reasons.  Firstly, there is a conformal factor of $r$ multiplying the
entire metric, and secondly, the three-dimensional transverse space has
the non-standard metric $dr^2 + \ft14 r^2\, d\Omega_2^2$.  This metric
suffers from a diverging curvature as $r$ approaches zero, since
the foliating 2-spheres are of the wrong radius to ``nest'' nicely
around the origin.  We see that it is natural to perform a coordinate
transformation in which we define a new radial coordinate $\rho =
\ft14 r^2$.  In terms of this, the metric (\ref{d41}) becomes
%%%%%
\be
ds_4^2 = -\Big(\fft1{\rho}\, H_e H_m K_w\Big)^{-\ft12} \, dt^2 +
\Big(\fft1{\rho}\, H_e H_m K_w\Big)^{\ft12}\, (d\rho^2 + \rho^2\,
d\Omega_2^2) \ .\label{d42}
\ee
%%%%%
Note that in terms of the new radial coordinate, the original harmonic
functions $H_e$, $H_m$ and $K_w$ given in (\ref{d5harm}) become
%%%%%
\be
H_e =1 + \fft{q}{\rho}\ ,\qquad H_m=1 + \fft{p}{\rho}\ ,\qquad
K_w = 1 + \fft{Q_w}{\rho}\ ,\label{d4harm}
\ee
%%%%%
which are harmonic in the new 3-dimensional transverse space.
Thus we see that the effect of performing a Hopf reduction on a
3-charge black hole in $D=5$ is to give the solution
%%%%%
\be
ds_4^2 = -\Big(K_N H_e H_m K_w\Big)^{-\ft12} \, dt^2 +
\Big(K_N H_e H_m K_w\Big)^{\ft12}\, (d\rho^2 + \rho^2\,
d\Omega_2^2) \ ,\label{d43}
\ee
%%%%%
which is precisely of the form of a standard 4-charge black hole,
except that the fourth harmonic function $K_N=1/\rho$, which has
charge equal to 1, is lacking a constant term.  Of course if we
consider the near-horizon limit where $\rho$ approaches zero, the
absence of the constant term becomes immaterial.  In fact, it is also
possible to introduce a constant term in the harmonic function $K_N$,
by performing an appropriate U-duality transformation \cite{hyun,%
Boonstra1,cllpst,bb}.   Thus we see that the
Hopf dimensional reduction provides a natural isentropic mapping
between a 3-charge black hole in $D=5$ and a 4-charge black hole in $D=4$.
The Hopf reduction preserves the supersymmetry of the solution.
Although the four-dimensional black hole involves 4 charges, it
still preserves the same fraction $1/8$ of the supersymmetry as does
the 3-charge black hole in $D=5$. (With the opposite orientation for
the Hopf reduction, all the supersymmetry is broken, which is
consistent with the fact that 4-charge black holes in $D=4$ also occur
with no preserved supersymmetry, if the sign of the fourth charge is
reversed \cite{lpmult,khuri}.)

The crucial feature in the above discussion is that the foliating
spheres in the transverse space are 3-dimensional.  One can perform a
Hopf reduction on the $U(1)$ fibres of any odd-dimensional sphere
$S^{2n+1}$, but in general the base space will then be $CP^n$.
However, it is only in the special case $n=1$ that the transverse
space after the Hopf reduction can become flat, with no singularity
at the origin.  This is because a metric of the form $d\rho^2 +
\rho^2\, d\Sigma_n^2$, where $d\Sigma_n^2$ is the metric on $CP^n$, is
non-singular at the origin only if $d\Sigma_n^2$ is a metric on the
unit sphere.  Only for $n=1$ is $CP^n$ isomorphic to a sphere
($CP^1\sim S^2$).

     In fact, we can apply the Hopf reduction to any $N$-charge $p$-brane
solution whose transverse space is 4-dimensional, and thereby obtain
an $(N+1)$-charge $p$-brane solution in one dimension less, with the
strength of the extra charge being 1.  (As in the previous example of
$D=5$ black holes, a further U-duality transformation is needed in
order to introduce a constant term in the associated harmonic
function.)  Note that the procedure can equally well be applied to the
case of non-extremal $p$-branes.

   Let us consider a general isotropic $N$-charge non-extremal $p$-brane
solution in $D+1$ dimensions, where the transverse space has
dimension 4.  The metric will have the form \cite{dlpblack}
%%%%%
\be
ds_{\sst{D+1}}^2 = \prod_{i=1}^N H_i^{-2/(D-1)}\, \Big( -e^{2f}\, dt^2
+ d\vec x\cdot d\vec x \Big) + \prod_{i=1}^N H_i^{(D-3)/(D-1)}\,
\Big( e^{-2f}\, dr^2 + r^2 \, d\Omega_3^2 \Big)\ ,
\ee
%%%%%
where the ``harmonic'' functions take the form $H_i = 1 + 4 k\,
\sinh^2\mu_i\, r^{-2}$, and the ``blackening function'' $f$ is given by
$e^{2f} = 1-k\, r^{-2}$. (Full details of the solutions, and the
relation of the charges and the mass to the parameters $\mu_i$ and $k$
can be found in \cite{dlpblack}.)  Following the same steps as we
described above for the reduction of the 3-charge $D=5$ black hole,
we find that the Hopf-reduced solution in $D$ dimensions has the
metric
%%%%%
\be
ds_{\sst{D}}^2 = \prod_{i=1}^{N+1} H_i^{-1/(D-2)}\, \Big( -e^{2f}\, dt^2
+ d\vec x\cdot d\vec x \Big) + \prod_{i=1}^{N+1} H_i^{(D-3)/(D-2)}\,
\Big( e^{-2f}\, d\rho^2 + \rho^2 \, d\Omega_2^2 \Big)\ ,
\ee
%%%%%
where $\rho=\ft14 r^2$, and we now have $e^{2f} = 1 -k/(4\rho)$ and
%%%%%
\be
H_i= 1+\fft{k}{\rho}\, \sinh^2\mu_i \ ,\quad (1\le i\le N)\ ,\qquad
H_{\sst{N+1}} = \fft{1}{\rho}\ .\label{nplus1}
\ee
%%%%%
As before, we can then perform a suitable U-duality transformation on
this solution in order to introduce a constant term in the new
harmonic function $H_{\sst{N+1}}$ \cite{hyun,Boonstra1,cllpst,bb}.
The metric (\ref{nplus1}) is then precisely of the form of an
$(N+1)$-charge $p$-brane in $D$ dimensions, with unit strength for the
extra harmonic function.  The discussion extends to any intersection
of $p$-branes, NUTs and waves where the overall transverse space is
4-dimensional.

In general, the near-horizon limit of an $N$-charge extremal $p$-brane
is singular. It follows in these cases that the Hopf reduction will
give rise to a singular dilaton in the lower dimension.  Thus the
supersymmetry will be further broken in such cases.

An application of our discussion of Hopf reductions for 4-dimensional
transverse spaces is to 5-branes in $D=10$.  There are NS-NS 5-branes,
and R-R D5-branes.  For the case of NS-NS 5-branes, Hopf reduction on
the $U(1)$ fibres of $S^3$ gives rise to 2-charge harmonic 5-branes in
$D=9$.  Performing a T-duality transformation, the solution can be
oxidised to $D=10$, when it again becomes an NS-NS single-charge
5-brane, this time with unit charge.  At the same time, the foliating
3-spheres become lens spaces $S^3/Z_n$, where $n$ is the magnetic
charge of the original 5-brane.  Thus a single ({\it i.e.}\ unit
charge) 5-brane is invariant under the Hopf T-duality.  The picture is
more complicated for the case of R-R 5-branes.  The Hopf reduction to
$D=9$ gives the same metric form as in the NS-NS case, but after
T-duality it oxidises back to the intersection of an NS-NS 5-brane and
a D6-brane.  In this case, the original 3-sphere is untwisted to
$S^2\times S^1$ (but with a non-direct-product metric, since there
will be different harmonic functions multiplying the two factors).

\bigskip\bigskip
\noindent{\Large{\bf Acknowledgment}}
\bigskip

    H.L. and C.N.P. are grateful to CERN for hospitality.  We thank Sudipta 
Mukherji for useful discussions.


\bigskip\bigskip

\newpage

\addcontentsline{toc}{part}{References}

\begin{thebibliography}{99}

\bibitem{DGT}
M.J. Duff, G.W. Gibbons and P.K. Townsend,
{\sl Macroscopic superstrings as interpolating solitons},
Phys. Lett. {\bf B332} (1994) 321, hep-th/9405124.

\bibitem{GHT}
G.W. Gibbons, G.T. Horowitz and P.K. Townsend,
{\sl Higher-dimensional resolution of dilatonic black hole
singularities},
Class. Quant. Grav. {\bf 12} (1995) 297, hep-th/9410073.

\bm{lublack} M.J. Duff and J.X. Lu, {\sl Black and super $p$-branes in
diverse dimensions}, Nucl. Phys. {\bf B416} (1994) 301, hep-th/9306052.

\bibitem{DKL}
M.J. Duff, R.R. Khuri and J.X. Lu,
{\sl String Solitons},
Phys. Rep. {\bf 259} (1995) 213, hep-th/9412184.

\bm{Rahmfeld} M.J. Duff, S. Ferrara, R. Khuri and J. Rahmfeld,
{\sl Supersymmetry and dual string solitons}, Phys. Lett. {\bf B356}
(1995) 479, hep-th/9506057.

\bm{DLPhet} M.J. Duff, H. L\"u, C.N. Pope, {\sl Heterotic phase
transitions and singularities of the gauge dyonic string}, hep-th/9603037.

\bm{DLPgauge} M.J. Duff, James T. Liu, H. L\"u, C.N. Pope, {\sl Gauge
dyonic strings and their global limit}, hep-th/9711089.

\bm{Boonstra1} H.J. Boonstra, B. Peeters and K. Skenderis, {\sl
Duality and asymptotic geometries}, Phys. Lett. {\bf B411} (1997) 59,
hep-th/9706192.

\bm{Boonstra2} H.J. Boonstra, B. Peeters and K. Skenderis, {\sl Brane
intersections, anti-de Sitter spacetimes and dual superconformal
theories}, hep-th/9803231.

\bibitem{Horowitz2}
G.T. Horowitz, J. Maldacena and A. Strominger,
{\sl Nonextremal black hole microstates and U duality},
hep-th/9603109.

\bibitem{Horowitz3}
G.T. Horowitz and A. Strominger,
{\sl Counting states of near extremal black holes},
hep-th/9602051.

\bibitem{Maldacena3}
J. Maldacena and A. Strominger,
{\sl AdS$_{3}$ black holes and a stringy exclusion principle},
hep-th/9804085.

\bm{nk2} N. Kaloper, {\sl Entropy count for extremal three-dimensional 
black strings}, hep-th/9804062.

\bm{callan} C. Callan and J. Maldacena, {\sl D-brane approach to black
hole quantum mechanics}, hep-th/9602043.

\bm{Tseytlin1} A.A. Tseytlin, {\sl Extreme dyonic black holes in string
theory},  Mod. Phys. Lett. {\bf A11} (1996) 689, hep-th/9601177.

\bm{Cvetic1} M. Cvetic and A.A. Tseytlin, {\sl General class of BPS
saturated black holes as exact superstring solutions}, Phys. Lett.
{\bf B366} (1996) 95, hep-th/9510097.

\bm{Cvetic2} M. Cvetic and A.A. Tseytlin, {\sl Solitonic strings and
BPS saturated dyonic black holes}, Phys. Rev. {\bf D53} (1996) 5619,
hep-th/9512031.

\bm{lpmult} H. L\"u and C.N. Pope, {\sl Multi-scalar $p$-brane
solitons}, Int. J. Mod. Phys. {\bf A12} (1997) 437, hep-th/9512153.

\bm{Vafa} A. Strominger and C. Vafa, {\sl Microscopic origin of the
Beckenstein-Hawking entropy}, hep-th/9601029

\bm{DBISSS} D. Birmingham, I. Sachs and S. Sen, {\sl Entropy of
three-dimensional black holes in string theory}, hep-th/9801019.

\bm{Maldacena} J. Maldacena, {\sl The large $N$ limit of
  superconformal field theories and supergravity}, hep-th/9711200.

\bm{behrndt} K. Behrndt, I. Brunner and I. Gaide, {\sl Entropy and
conformal field theories of AdS$_{3}$ models}, hep-th/9804159.

\bm{evans} J. M. Evans, M. R. Gaberdiel and M. J. Perry, {\sl The
no-ghost theorem for AdS$_{3}$ and the stringy exclusion principle},
hep-th/9806024.

\bm{andrianopoli} L. Andrianopoli, R. D'Auria, S. Ferrara and M.A. Lledo
{\sl Horizon geometry, duality and fixed scalars in six dimensions},
hep-th/9802147.

\bm{navarro} J. Navarro-Salas and P. Navarro, {\sl A note on
AdS$_{3}$ and boundary conformal field theory}, hep-th/9807019.

\bm{lee} H.W. Lee, N.J. Kim and Y.S. Myung, {\sl Dilaton test of
connection between AdS$_{3} \times S^{3}$ and $5D$ black hole},
hep-th/9805050.

\bm{andreas} B. Andreas, G. Curiol and D. Lust, {\sl The
Neveu-Schwarz fivebrane and its dual geometries}, hep-th/9807008.

\bm{deger} S. Deger, A. Kaya, E. Sezgin and P. Sundell,
{\sl Spectrum of $D=6$ $N=4b$ supergravity on AdS$_{3}\times
S^{3}$}, hep-th/9804166.

\bm{benakli} K. Benakli, {\sl Heterotic string theory on anti-de Sitter
spaces}, hep-th/9805099.

\bm{Boer} J. de Boer, {\sl Six-dimensional supergravity on
$S^{3}\times AdS_{3}$}, hep-th/9806104.

\bm{Vafa2} C. Vafa, {\sl Puzzles at large N}, hep-th/9804172.
\bibitem{Boonstra}
H.J. Boonstra, B. Peters, K. Skenderis,
{\sl Branes and anti de Sitter space-time},
hep-th/9801076.

\bibitem{CKKTV}
P. Claus, R. Kallosh, J. Kumar, P.K. Townsend and A. Van
Proeyen, {\sl Supergravity and the large $N$ limit of theories with
sixteen supercharges},
hep-th/9801206.

\bm{Ortiz} M. Banados and M. Ortiz, {\sl The central charge in three
dimensional anti-de Sitter space}, hep-th/9806087.

\bm{giveon} A. Giveon, D. Kutasov and N. Seiberg, {\sl Comments on
string theory on AdS$_{3}$}, hep-th/9806194.

\bm{martinec} E. Martinec, {\sl Matrix models of AdS gravity}, hep-th/9804111

\bibitem{DLP2} M.J. Duff, H. L\"u and C.N. Pope, {\sl AdS$_{5} \times
    S^{5}$ untwisted}, hep-th/9803061.

\bibitem{swos}
M.J. Duff, H. L\"u and C.N. Pope,
{\sl Supersymmetry without supersymmetry},
Phys. Lett. {\bf B409} (1997) 136, hep-th/9704186.

\bibitem{DLPS} M.J. Duff, H. L\"u, C.N. Pope and E. Sezgin, {\sl
    Supermembranes with fewer supersymmetries}, Phys. Lett. {\bf B371}
  (1996) 206, hep-th/9511162.

\bm{ferrara} S. Ferrara, A. Kehagias, H. Partouche and A. Zaffaroni,
{\sl Membranes and fivebranes with lower supersymmetry and their AdS
duals}, hep-th/9803109

\bm{ccdfft} L. Castellani, A. Ceresole, R. D'Auria, S. Ferrara,
P. Fr\'e and M. Trigiante, {\sl $G/H$ M-branes and AdS$_{p+2}$ geometries},
hep-th/9803039.

\bibitem{kachru}
S. Kachru and E. Silverstein,
{\sl $4d$ conformal field theories and strings on orbifolds},
hep-th/9802183.

\bm{Oz} Y. Oz and J. Terning, {\sl Orbifolds of AdS$(5) \times S^{5}$
and 4-D conformal field theories}, hep-th/9803167.

\bm{klebanov} I. Klebanov and E. Witten, {\sl Superconformal field
theory on three-branes at a Calabi-Yau singularity}, hep-th/9807080.

\bm{llpt} I.V. Lavrinenko, H. L\"u, C.N. Pope and T.A Tran, {\sl
  U-duality as general coordinate transformations, and spacetime
  geometry}, hep-th/9807006.

\bm{schwarz} J.H. Schwarz, {\sl An $SL(2,\Z)$ multiplet of
  superstrings}, Phys. Lett. {\bf B360} (1995) 13; Erratum-ibid.
{\bf B364} (1995) 252, hep-th/9508143.

\bm{lag} E. Alvarez, L. Alvarez-Gaume, J.L.F. Barbon and Y. Lozano,
{\sl Some global aspects of duality in string theory}, Nucl. Phys.
{\bf B415} (1994) 71, hep-th/9309039.

\bm{lpsol} H. L\"u and C.N. Pope, {\sl $p$-brane solitons in maximal
supergravities}, Nucl. Phys. {\bf B465} (1996) 127, hep-th/9512012.

\bm{cjlp1} E. Cremmer, B. Julia, H. L\"u and C.N. Pope, {\sl
  Dualisation of dualities}, Nucl. Phys. {\bf B523} (1998) 73,
  hep-th/9710119.

\bm{buscher} T.H. Buscher, {\sl A symmetry of the string background
 field equations}, Phys. Lett. {\bf B194} (1987) 59.

\bm{np} B.E.W. Nilsson and C.N. Pope, {\sl Hopf fibration of
$D=11$ supergravity}, Class. Quantum Grav. {\bf 1} (1984) 499.

\bm{horowitz} G.T. Horowitz and D.L. Welch, {\sl Exact
  three-dimensional black holes in string theory},
  Phys. Rev. Lett. {\bf 71} (1993) 328, hep-th/9302126.

\bm{nk1} N. Kaloper, {\sl Miens of the three-dimensional black
hole}, Phys. Rev. {\bf D48} (1993) 2598, hep-th/9303007.

\bm{lpt} H. L\"u, C.N. Pope and P.K. Townsend, {\sl Domain walls from
  anti-de Sitter spacetime}, Phys. Lett. {\bf B391} (1997) 39,
hep-th/9607164.

\bm{fy} Y. Fujii and K. Yamagishi, {\sl Killing spinors on spheres and
  hyperbolic manifolds}, J. Math. Phys. {\bf 27} (1986) 979.

\bm{lpr} H. L\"u, C.N. Pope and J. Rahmfeld, {\sl A construction of
  Killing spinors on $S^n$}, hep-th/9805151.

\bm{khuri} R.R. Khuri and T. Ortin, {\sl A non-supersymmetric dyonic
  extreme Reissner-Nordstrom black hole}, Phys. Lett. {\bf B373}
  (1996) 56, hep-th/9512178.

\bm{classp} H. L\"u, C.N. Pope, T.A. Tran and K.-W. Xu, {\sl
Classification of $p$-branes, NUTs, waves and intersections},
Nucl. Phys. {\bf B511} (1998) 98, hep-th/9708055.

\bm{hyun}S. Hyun, {\sl U-duality between three and higher dimensional
black holes}, hep-th/9704005.

\bm{cllpst} E. Cremmer, I.V. Lavrinenko, H. L\"u, C.N. Pope,
K.S. Stelle and T.A. Tran, {\sl Euclidean-signature supergravities,
dualities and instantons}, to appear in Nucl. Phys. {\bf B}, hep-th/9803259.

\bm{bb} E. Bergshoeff and K. Behrndt, {\sl D-instantons and asymptotic
geometries}, Class. Quant. Grav. {\bf 15} (1998) 1801, hep-th/9803090.

\bm{cjlp2} E. Cremmer, B. Julia, H. L\"u and C.N. Pope, {\sl
  Dualisation of dualities II: Twisted self-duality of doubled fields
  and superdualities}, hep-th/9806106.

\bm{duffrahm} M.J. Duff and J. Rahmfeld, {\sl Massive string states as
extreme black holes}, Phys. Lett. {\bf B345} (1995) 441, hep-th/9406105.

\bm{trombone} E. Cremmer, H. L\"u, C.N. Pope and K.S. Stelle, {\sl
  Spectrum-generating symmetries for BPS solitons}, Nucl. Phys. {\bf
  B520} (1998) 132, hep-th/9707207.

\bm{lr} J.X. Lu and S. Roy, {\sl U-duality $p$-branes in diverse
dimensions}, hep-th/9805180.

\bm{blps} M.S. Bremer, H. L\"u, C.N. Pope and K.S. Stelle, {\sl Dirac
  quantisation conditions and Kaluza-Klein reduction}, to appear in
Nucl. Phys. {\bf B}, hep-th/9710244.

\bm{lpsorbit} H. L\"u, C.N. Pope and K.S. Stelle, {\sl Multiplet
  structures of BPS solitons}, Class. Quant. Grav. {\bf 15} (1998)
  537, hep-th/9708109.

\bm{fg} S. Ferrara and M. Gunaydin, {\sl Orbits of exceptional groups,
duality and BPS states in string theory}, Int. J. Mod. Phys. {\bf
A13} (1998) 2075, hep-th/9708025.

\bm{dlpblack} M.J. Duff, H. L\"u and C.N. Pope, {\sl The black branes
  of M-theory}, Phys. Lett. {\bf B382} (1996) 73, hep-th/9604052.

\bm{kiritsis} E. Kiritsis, {\sl Exact duality symmetries in CFT and 
string theory}, Nucl. Phys. {\bf B 405} (1993) 109, hep-th/9302033.

\bm{antoniadis} I. Antoniadis, C. Bachas and A. Sagnotti, {\sl Gauged 
supergravity vacua in string theory}, Phys. Lett. {\bf B 235} (1990) 
255.

\end{thebibliography}
\end{document}